\documentclass[final,10pt,times,3p]{elsarticle}
\usepackage{graphicx, amsfonts, amsmath, amssymb}
\usepackage[utf8]{inputenc}
\usepackage{adjustbox}
\usepackage{array}
\usepackage{color,empheq, colortbl}
\usepackage{xcolor}
\usepackage{float}
\usepackage[colorlinks = true,
            linkcolor = blue,
            urlcolor  = magenta,
            citecolor = red,
            anchorcolor = green]{hyperref}
\newcolumntype{L}[1]{>{\raggedright\let\newline\\\arraybackslash\hspace{0pt}}m{#1}}
\newcolumntype{C}[1]{>{\centering\let\newline\\\arraybackslash\hspace{0pt}}m{#1}}
\newcolumntype{R}[1]{>{\raggedleft\let\newline\\\arraybackslash\hspace{0pt}}m{#1}}
\newcolumntype{M}[1]{>{\centering\arraybackslash}m{#1}}
\newcolumntype{N}{@{}m{0pt}@{}}
\usepackage{multirow}
\newcommand{\bra}[1]{\left\langle{#1}\right\vert}
\newcommand{\ket}[1]{\left\vert{#1}\right\rangle}
\definecolor{LightCyan}{RGB}{204,255,255}
\definecolor{LightGreen}{RGB}{153,255,204}
\definecolor{Yellow}{RGB}{255,255,204}
\definecolor{LightRed}{RGB}{255,204,255}
\begin{document}
%opening
\title{How efficient is transport of quantum cargo through multiple highways?}
%Sending quantum Cargo through multiple highways are more beneficial than using 
%\author{Saptarshi Roy}
%\email{saptarshiroy@hri.res.in}
%\affiliation{Harish-Chandra Research Institute, HBNI, Chhatnag Road, Jhunsi, Allahabad 211019, India}
%\author{Tamoghna Das}
%\email{tamoghna@hri.res.in}
%\affiliation{Harish-Chandra Research Institute, HBNI, Chhatnag Road, Jhunsi, Allahabad 211019, India}
%\author{Debmalya Das}
%\email{debmalyadas@hri.res.in}
%\affiliation{Harish-Chandra Research Institute, HBNI, Chhatnag Road, Jhunsi, Allahabad 211019, India}
%\author{Aditi Sen (De)}
%\email{aditi@hri.res.in}
%\affiliation{Harish-Chandra Research Institute, HBNI, Chhatnag Road, Jhunsi, Allahabad 211019, India}
%\author{Ujjwal Sen}
%\email{ujjwal@hri.res.in}
%\affiliation{Harish-Chandra Research Institute, HBNI, Chhatnag Road, Jhunsi, Allahabad 211019, India}
%\author{Saptarshi Roy, Tamoghna Das, Debmalya Das, Aditi Sen(De) and Ujjwal Sen}
\author{Saptarshi Roy\(^1\), Tamoghna Das\(^{1,2}\), Debmalya Das\(^{1,3,4}\), Aditi Sen(De)\(^1\), Ujjwal Sen\(^1\)}

\address{\(^1\)Harish-Chandra Research Institute, HBNI, Chhatnag Road, Jhunsi, Allahabad 211 019, India.}
\address{\(^2\)International Centre for Theory of Quantum Technologies,  University of Gda\'{n}sk, 80-952 Gda\'{n}sk, Poland.}
\address{\(^3\)Department of Physics and Astronomy, University of Rochester, Rochester, NY 14627, USA.}
\address{\(^4\)Center for Coherence and Quantum Optics, University of Rochester, Rochester, NY 14627, USA.}

%\shortauthor{S. Roy \etal}

%\address{Harish-Chandra Research Institute, HBNI, Chhatnag Road, Jhunsi, Allahabad 211019, India}

%%%%%%%%%%%%%%%%%%%%%%%%%%%%%%%%%%%%%%%%%%%%%%%%%%%%%%%%%%%%%%%%%%%%%%%
\begin{abstract}
%Quantum states can be efficiently transferred over a long distance if the
%entire quantum channel can be divided into several small blocks. 
%\textcolor{red}{Quantum states can be efficiently transferred with the help of finite amount of classical communication if the sender and the receiver share an entangled state via the quantum teleportation protocol.  Here we introduce one or more intermediate parties which are required if the sender is compelled to leave the protocol before finishing the task while multiple intermediate parties can be useful for completing the protocol in the pre-decided time. Specifically, 
% we consider a scenario in which two copies of
%a multiparty state are shared} 
Quantum states can be efficiently transferred over a long distance if the
entire quantum channel can be divided into several small blocks. We
consider a scenario in which each block consists of two copies of
a multiparty state -- one is used for distributing an arbitrary quantum state to
multiple parties while the other channel is required to concentrate it back to
a single party. Both in noiseless and local noisy scenarios,  we find one-shot
quantum capacities of these channels in terms of fidelity, when the
initial shared states in each block  are the generalized
Greenberger-Horne-Zeilinger and the generalized W states. We also consider a
situation where optimal local measurements transform multipartite states
to bipartite ones which can then be used as single-path channels for quantum
state transmission in each segment. We show that in some parameter ranges,
the former protocol provides strictly better fidelities than that of the latter,
thereby establishing the importance of  distributing and concentrating
arbitrary quantum states via multipartite entangled states  in long distance quantum
communication, over the local measurement based protocol. Moreover, we show that in presence of bit flip or bit-phase flip noise, shared generalized Greenberger-Horne-Zeilinger states  possess an inherent noise detection and correction mechanism, leading to the same fidelity as in the noiseless case. We consider further noise models also, which do not enjoy the same mechanism.
In addition to the fidelity based advantages, the multipath scheme is shown to be useful when one considers a  situation in which the completion of the teleportation needs to be delayed. We also find the efficiencies of a quantum channel when a quantum state is transferred over long distances and the entire channel is divided into several small blocks.

%Quantum state transfer protocol or quantum internet is proposed over long distance, divided into several blocks where each block consists of a multiparty entangled states  either distributing the information content of the unknown qubit through various edges of the multiparty entangled state or concentrating it back via the same state. 
%The fidelities obtained in the multiparty scheme are then compared to those of the entire process, consisting of several blocks of bipartite states
%obtained via optimal measurements performed locally on  qubits of each block of a multiparty state -- the former can be called multipath quantum repeater while the later is the measurement-based single path repeater. We compute  these fidelities for shared generalised GHZ and generalised W states for both the schemes.
%We report a distinct advantage of generalized W states when we opt for the multiparty  protocol instead of the  measurement-based  scheme in a noiseless scenario, which is not the case for the shared generalized GHZ state.
%However, when noise acts on one of the qubits in each block, for some parameter regions of both gGHZ and gW states, the multipath quantum repeater is better than the measurement-based single path ones. Moreover, in case of bit flip noise and when the shared state is gGHZ state, we show that the proposed scheme can detect the noise and correct it only in the multipath case.
\end{abstract}

\maketitle
%%%%%%%%%%%%%%%%%%%%%%%%%%%%%%%%%%%%%%%%%%%%%%%%%%%%%%%%%%%%%%%%%%%%%%
\section{Introduction}

From the dawn of civilization, communication between different persons and groups played a crucial role
in shaping human society. 
%Information communicated from one group to another during several critical times have 
%been pivotal in changing the course of history. 
With the emergence of modern science and technologies,
a new era in the transmission of information with and without security has been developed.  In both the cases, it has been  realized, both theoretically \cite{horodecki2009, BB84, DC, Crypto, bennett1993} and experimentally \cite{communication_exp, tele_review}, that quantum technologies can provide higher efficiencies  \cite{horodecki2009} than their classical counterparts.
%For example, quantum cryptography \cite{Crypto} turns out to be more secure than its classical version,
%the capacity of classical information transmission can be increased upto double with the help of quantum states \cite{DC}.
In most of the cases, the main ingredient that helps to acheive such advantage is the bipartite entanglement of a shared quantum state between the sender and the receiver \cite{horodecki2009}, (cf. \cite{error_correction, onewayQC, quantum_internet, state_tomo, Q_metrology, Phase_transit} for other quantum information processing tasks).

One of the most interesting consequences of bipartite entanglement
is the ability to transfer an arbitrary quantum state,  with the additional help of a finite amount of classical communication -- quantum teleportation~\cite{bennett1993}.
It is an indispensable feature of quantum mechanics, since to achieve this task by using unentangled states,  one requires infinite amount of classical communication between the sender, Alice, and the receiver, Bob (cf. \cite{horo_fid, bound_activation}).
Significantly, within few years of its discovery, it has been realized experimentally \cite{bennett1993, tele_review}, first by using photons \cite{tele_photon_exp,tele_photon_exp2, tele_contvar_exp, photon_loss_distance} and then in other physical substrates  \cite{tele_review, tele_ion_exp, tele_exp_coldatom, tele_nmr_exp, tele_supercond, tele_exp_light_matter, ren2017}, thereby establishing a new epoch in communication, which can potentially lead to a quantum internet \cite{quantum_internet}, parallel to the extremely useful internet already in use.

One of the avenues by which this breakthrough can become more prominent 
%Such an  important discovery can have high commercial value 
is by the involvement of multiple parties, which will be a step-forward towards  building a quantum network. Over the years, several quantum information transmission protocols with multipartite states  were proposed and some of them have also been demonstrated in experiments \cite{murao1998, muralidharan2016}.
On the other hand, if Alice and Bob are located over a long distance, it has been noticed that the task of teleporting an unknown quantum state from Alice to Bob by using a single entangled state is not the best
resort. 
%Since, noise may interfere  the capacity of transmission. 
For example,  in  case of photonic
systems with  polarization degrees of freedom,  photon loss  approximately becomes exponential with the length of the channel \cite{tele_photon_exp2, photon_loss_distance}, reducing the quantum capacity of a quantum channel.
As a remedy,  it was proposed that in a noisy environment, quantum state transfer over a long distance can be divided into  several small segments, in which entanglement distillation \cite{ent_purification} followed by a modified version of quantum teleportation, known as entanglement swapping \cite{ent_swapping} are performed to obtain quantum efficiency -- the entire process being called a quantum repeater (QR)  \cite{q_repeater}.

In this work,  we consider a multipath teleportation protocol where the quantum information is distributed through various pathways before concentrating it back to the desired sender. For some parameter ranges, such a protocol offer better fidelity compared to single path schemes both in the presence and absence of noise.
Having specified one advantage of using multiple pathways instead of a single one, let us now mention another advantage of the same, which however preassumes the existence of a quantum memory. In this case,  we consider a teleportation protocol in which Alice would like to teleport an
arbitrary state to Bob, but at a predetermined later time, at which she would not be able to be present at her
port. Therefore, the process has to be mediated by an intermediate party (Claire) who would complete the 
objective of the process. However, if the intermediate party is not trustworthy, she/he  might  
teleport the state before the predecided time. To reduce this possibility, Alice decides to resort to multiple
Claires, so that the process is completed if and only if all the mediators complete their actions. Therefore,
even if some Claires break the trust, Bob does not receive the state before the predecided time. More Claires 
would naturally imply a better assurance in terms of the completion of the protocol in the required time.
% Specifically, if one of the mediators breaks trust by performing the operations at her port at an earlier time, the process is not completed and it ends if and only if all the mediators complete their actions. 
The above two arguments explain the primary motivation behind considering a multipath teleportation scheme.

At this point, we want to mention that to distribute entanglement over large distance, there are two important schemes in the literature —- entanglement percolation \cite{per} and quantum repeaters \cite{q_repeater}. In the classical percolation of entanglement (CEP), non maximally entangled or less entangled states are transferred  to singlets or Greenberger Horne Zeilinger (GHZ) states with an optimal probability by local operations followed by classical communication and then entanglement swapping  leads to the distribution of entanglement among nodes situated in distant location. It was shown that depending on the lattice structure, if one initially performs certain operations which can transform lattice to an optimal setting, and then CEP can be applied, one can  enhance the probability of entanglement distribution. On the other hand, for repeaters, starting from many copies of a noisy entangled states, a maximally entangled state is created via distillation and after that, entanglement swapping is performed between adjacent nodes to create entangled states in distant parties.

\begin{figure}
\begin{center}
\includegraphics[width=\linewidth]{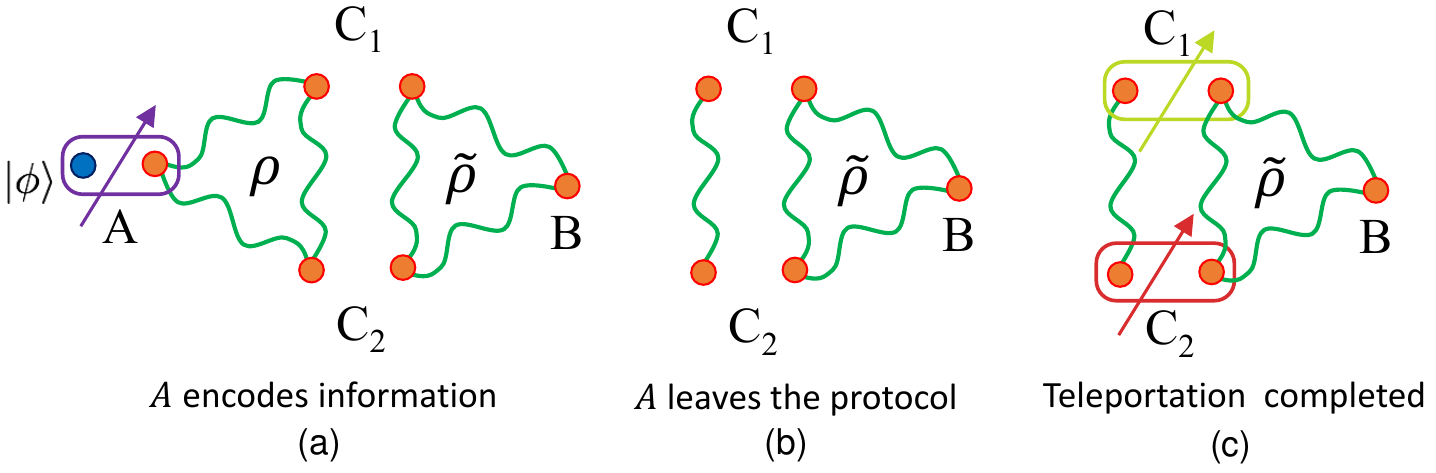}
\end{center}
\caption{\label{fig:TDTC} A schematic diagram of multipath quantum repeater.  It consists of teledistribution and teleconcentration parts. A sender, Alice, $A$, who posses a unknown quantum state to be teleported to a receiver, Bob, denoted by $B$ shares a multiparty state, \(\rho\), with Claires, \(C_1\) and \(C_2\), while $B$ is also connected with the same Claires via another multiparty state, \(\tilde{\rho}\).  $A$ and $B$ are  respectively called input and output ports while Claires are the auxiliary nodes. (a)- First step: $A$ performs a joint measurement on her part of the shared state and the state to be teleported. (b)- The scenario when $A$ has left the protocol after the joint measurement by $A$ and unitary operations by $C_1$ and $C_2$ on $\rho$. (c) Last step is   when $C_1$ and $C_2$ perform their measurements and complete the process.
%state. Alice shares an multipartite entangled state with the Claires who in turn share another
%similar state with Bob. The state to be teleported is coupled with the input port, with Alice,
%who performs a joint Bell measurement locally on them and communicates the result calssically to
%the Claires. The Claires accordingly apply suitable unitaries on their qubits, shared with Alice
%and then perform Bell measurements on the qubits, each of them have. The results are then 
%classically communicated to Bob, who again apply suitable unitaries on his qubit, to obtain
%the output state
}
\end{figure}

Inspired by recent experimental achievements which establish an entangled channel over
a few thousand kilometers \cite{photon_loss_distance} and  developments on QRs \cite{QRrecentpapers}, we consider 
a quantum state transmission protocol, which is divided into several small blocks, involving mutipartite states.
In particular, each  block  consists of two multipartite states, among which  one is used to transfer an arbitrary 
quantum state among multiple parties while the other one is for accumulating it back to a single one. We call this
process as the ``multipath quantum repeater" which has two components, ``teledistribution" and ``teleconcentration"
(TD-TC) (see Fig. \ref{fig:TDTC}). The port which possesses the initial state to be transferred can be referred as
an input port while the receiver who finally receives the unknown state is the output port, and other parties 
involved in this protocol can be called the auxiliary nodes. 
%At this point, there are two natural questions that one can ask:
To summarize, Alice achieves the delayed teleportation without she being available at her port but also ensuring the specific time of teleportation by:
\begin{enumerate}
\item Invoking a third party, Claire, between Alice and Bob who would mediate the teleportation process at the predecided time.
\item Involving two or more Claires which provides a better assurance of the perfect timing of the protocol even if some of the
intermediates do not abide  by the temporal guideline.
\end{enumerate}
  %\begin{enumerate}
%\item Invoking a third party in each block, i.e., employ a distribution and concentration scheme?
%\item What is the reason behind two or more Claires?
%\end{enumerate}    
%It is true that the main objective is to teleport an unknown quantum state from the sender, Alice to the receiver, Bob. Let us suppose a situation where Alice wants to teleport an unknown qubit to Bob at some later time when she is absent at her port . She can achieve this task by mediating the teleportation process through a third party, Claire, who is responsible for executing the tasks in a later predetermined time. To address the second question, we first notice that in the aforementioned case,  Claire teleports the quantum state to Bob at the mutually agreed time, solely rests on how trustworthy Claire is. The perfect timing of the protocol can be ensured by introducing multiple mediators. If one of the mediators breaks trust by performing the operations at her port at an earlier time, the process is not completed and it ends if and only if all the mediators complete their actions. This explains the primary motivation behind considering a multipath teleportation scheme. 
As we will report, the results obtained in this scenario also shows some other advantages of this process. Of course, there is an established fact that sharing entangled state in a smaller distance compared to a longer one, is always advantageous due to the presence of decoherence.

The multiple-path protocol can reduce to a bipartite linear-chain scenario if all the auxiliary parties excepting
one, perform optimal local measurements -- we refer to such scheme as ``local measurement-based single-path QR". 
Such a situation might be relevant when one of the Claires refuses to take part in the protocol
but assists it, in terms of the fidelity yield, to the best of her ability by optimal local measurements.

In this paper, we compare (one shot) quantum capacities, in terms of  fidelities, obtained in these two processes
and report that both  in  noiseless and noisy  scenarios,  TD-TC protocol can yield higher fidelity, than that of
the local measurement-based single-path QR, for a set of three-qubit states,
justifying the importance of multipartite states, in QR process.  Specifically, in a noiseless case, we 
demonstrate this advantage  when the shared state is the three-qubit generalized W (gW) \cite{dur2000} state for 
certain range of parameters although such benefit  is not observed for the family of  generalized
Greenberger-Horne-Zeillinger (gGHZ) states \cite{gGHZ}. Interestingly, in presence of local noise in one or in both the 
auxiliary ports, the gGHZ and the gW states, 
for some choices of parameters show a clear advantage of TD-TC scheme with multiple paths over the single-path one. Moreover, we find that in 
the TD-TC protocol, the gGHZ state possesses an inherent local noise detection and rectification mechanism for
a specific type of noisy environment. 
Other relevant studies in similar 
lines include distribution accompanied by error correction of quantum information \cite{mu1} and remote information
extraction during teledistribution \cite{mu2}.
For a quantitative analysis, a numerical simulation is performed which gives the percentages of shared initial 
states, having better fidelity in multipath protocol than the local measurement-based ones in presence of several
paradigmatic noise models \cite{preskil,Nielsen_chuang},  namely, bit flip, phase flip, bit-phase flip, amplitude
damping and phase damping channels.
By using recursion relation, we finally obtain  quantum capacities of a quantum channel over a large distance,
having arbitrary number of blocks, when the gGHZ and the  gW states are used in each block, performing multipath 
as well as single-path schemes. 
%In this article, we restrict ourselves to the case of two mediators for the ease of computations.
In case of the gGHZ states, we also show that more mediators lead to the same fidelity as obtained with two mediators, even in the presence of bit flip, phase flip, bit phase flip noises. 
 
In our considered protocol, each block is constructed out of single copies of multipartite states enabling a one-shot process. It bears a strong resemblance with quantum repeaters without the initial step of entanglement distillation. We proceed to work with the states at hand without attempting to enhance its entanglement. Specifically, by performing measurements in different nodes (entanglement swapping), we establish entangled connections between two distant nodes. The novelty of our work lies on the fact that the proposal involves entanglement distribution followed by entanglement concentration which have distinct advantages like time delayed teleportation, detection and rectification of certain noise, etc. as mentioned before. Furthermore, compared to the percolation scheme,  our method can be applied to obtain fidelities for some mixed states as well, specifically those states that suffer from various local noise models. Note however that the computations of fidelity as well as designing the measurement-based protocol become harder both the noisy case and for initial shared mixed states.

The paper is arranged as follows. Sec. \ref{sec:dis_con} is devoted to a more detailed discussion of 
TD-TC protocol involving multipartite entangled states, between 
the source and the receiver. In Sec. \ref{sec:dis_conc_perform}, we discuss  the entire protocol in details for two families of three-qubit  pure states, namely the generalized GHZ and the generalized W states belonging to
two inequivalent SLOCC classes of three-qubit  states, as quantum channels in a multipath protocol while optimal fidelities in local measurement-based case are also 
 derived for the similar families of multiparty quantum states in Sec. \ref{sec:linearchain} for comparison.
Sec. \ref{sec:noise_tele} deals with the case when one of the parties are affected by different kinds of local noise. Before conclusion (Sec. \ref{sec:conclu}), we discuss the capacities of quantum state transfer when the entire distance is divided into several small blocks (see Sec. \ref{Sec:multi_block}).

%%%%%%%%%%%%%%%%%%%%%%%%%%%%%%%%%%%%%%%%%%%%%%%%%%%%%%%%%%%%%%%%%%%%%%%%%%%%%%%%%%%
\section{Multipath Quantum Repeater: Setting the stage}
\label{sec:dis_con}
%%%%%%%%%%%%%%%%%%%%%%%%%%%%%%%%%%%%%%%%%%%%%%%%%%%%%%%%%%%%%%%%%%%%%%%%%%%%%%%%%%%

In this section, we will discuss a communication protocol where a sender and a receiver are connected via multiple channels.
%the teledistribution and teleconcentration process \cite{murao2001}
%as shown in Fig. \ref{TDTC}. 
Alice, $A$, intends to send an unknown quantum state, $|\psi\rangle$, given by
\begin{equation}\label{eq:input}
 \ket{\psi}=a\ket{0}+b\ket{1},
\end{equation}
with $a$ and $b$ being complex numbers, satisfying $|a|^2 + |b|^2 = 1$, to Bob, $B$,
with the help of $N-1$ Claires, $C_1, C_2, \ldots, C_{N-1}$. 
Two multiparty quantum states, $\rho_N$, and $\tilde{\rho}_N$ are shared between Alice and  Claires, and  between  Claires and the Bob respectively (See Fig. \ref{fig:TDTC} for illustration with $N=3$). 
The protocol consists of the following steps: 

\begin{enumerate}
\label{enu1}
\item[Step 1.]  Alice initiates the process by performing a joint measurement, $M_A$, on the unknown input state $|\psi\rangle$, and her part of  the quantum state $\rho_N$. 

\item[Step 2.] Alice communicates her measurement outcome classically to the Claires,
$C_1,C_2, \ldots, C_{N-1}$, post which each of them performs local unitary operations,  $\{U^j_{C_i}\}, ~ i = 1,\ldots, N-1, ~ j=1, \ldots d$,  on their respective parts of the shared state $\rho_N$, where $d$ is the number of elements in the measurement basis. Alice may now leave the protocol.

\item[Step 3.] The $i^{\text{th}}$ Claire  performs a  measurement $\{M_{C_i}\}$, jointly on her part of the  rotated post-measured state of $\rho_N$ and her part of 
the shared $\tilde{\rho}_N$, at some later time which is predecided.

\item[Step 4.] Based on the measurement results communicated by the Claires,  Bob  rotates his part of the quantum state with unitary operators, $\{U_B^k\}$. We refer to such a protocol as ``multipath QR", where  Steps 1 and  2 are parts of TD while Steps 3 and  4 together constitutes the TC  (for the same, (cf. \cite{murao2001})).
\end{enumerate}

Let us suppose that at the end of the protocol, Bob obtains the state $\rho_{\psi}$. The fidelity under TD -TC is then defined as
\begin{equation}\label{eq:fidel1}
\tilde{\cal F}^{DC} = \int \langle \psi | \rho_{\psi}|\psi\rangle d\psi,
\end{equation}
which depends on the choice of measurements, $M_A, \{M_{C_i}\}_{i=1}^{N-1}$ at the Alice's and $N-1$ Claires' nodes respectively and also on the unitary rotations $\{U^j_{C_i}\}$ and $\{U_B^k\}$. 
%and it gives us a measure of how closely the unknown state $\ket{\psi}$,   is teleported at Bob's end. 
%
%It is clear that very poor choices of measurement strategies and unitary operators  may lead
%to a low teleportation fidelity. 
Hence, the optimal fidelity of multipath QR by shared multiparty quantum channels, $\rho_N$ and     
$\tilde{\rho}_N$,   is obtained by 
%to find out the usefulness of the shared multiparty quantum states
%$\rho_N$ and $\tilde{\rho}_N$,  one has to optimize
 maximizing Eq. (\ref{eq:fidel1}), over all measurement strategies and unitary operators, given by
\begin{equation}\label{eq:fidel}
{\cal F}^{DC}(\rho_N, \tilde{\rho}_N) = \max_{M_A, \lbrace M_{C_i}\rbrace, \{U^j_{C_i}\}, \{U_B^k\}}\int \langle \psi | \rho_{\psi}|\psi\rangle d\psi.
\end{equation}
Note  that  if a sender and a receiver share a
quantum state having vanishing entanglement, the teleportation fidelity of sending a qubit can not go beyond $\frac 23$, with the help of classical communication, while the fidelity reduces to $\frac 12$, when there is no classical communication allowed  between them. In this paper, whenever we encounter a product state, shared between Alice and  Claires or between the Claires and Bob, we put the value of fidelity to be $\frac 23$. Note that a pure quantum state between two parties, if without entanglement, can only be product.%, having no classical correlation.

%The other utility of using distribution and concentration 
%channels in teleportation is effecting a desired delay in the completion of the process after
%Alice has started with it. 
Before presenting the results, let us discuss the scenario where multipath QR can be useful. The advantages of QR is well known and hence we only concentrate on the benefit of  TD-TC protocol. 
Apart from the advantages obtained in terms of fidelity, we discuss the importance of the auxiliary nodes in the protocol and then the benefit of several nodes instead of a single one. First of all, consider a scenario where 
%To transfer an arbitrary quantum state, teledistribution and teleconcentration can be required if
 Alice wants to send an unknown quantum state at some predecided later time, when Alice is not available and Claires are ready to help Alice. In particular, Claires' presence is essential when 
% One of the goals of this protocol is to send the arbitrary quantum state $|\psi\rangle$, at some predecided later time or introducing a desired delay in the completion of the process after Alice starts it. 
%It follows from the fact that after Alice classically communicates her outcome and signals all the intermediate parties to carry out the protocol, all of them must perform their measurements and communicate their outcomes classically
%to the next party for the teleportation to be a success.
% It is legitimate to ask that for such a situation Alice should have started the protocol at that
% particular time she wants to send the quantum state to Bob. But suppose 
 Alice needs to leave the laboratory at some earlier time, or the location of her laboratory is in such a place that she could  not be present at the time of transmission. For example, such a situation can arise if  Alice is a part of secret agency and she
has to perform her measurement as soon as she gets the unknown state, any delay on her part
 can cause unnecessary threat to her as well as to the protocol.
 %We will show in this paper that the introduction of many Claires, as intermediaries, can be useful.
Secondly, it can also be argued that if the protocol involve a single Claires, and  if
 she is  ``uncooperative" or is compromised by any third party (enemy),  there is a possibility of a 
 measurement and a feed forward procedure, resulting in  information leakage before the time.  We will show that the introduction of many Claires, not only helps Alice-Bob to overcome Alice's constraints, but also leads to a better fidelity compared to a bipartite scenario. 
%In presence of many Claires, security comes due to the fact that not all of them can be untrustworthy or %traced and attacked simultaneously by a group of common enemy as they are at different locations.
In this paper,  we also assume that the Claires are not allowed to communicate between each other even classically, 
and classical communication is only between (Alice, $C_i$) ($i = 1,2,\ldots N$) and between ($C_i$, Bob) pairs.
%and only Alice can signal them, when the proper time comes.

Our aim here is to analyze the performance of the shared quantum channels, both noiseless and noisy,  in terms of  fidelity
of the TD-TC protocol defined in Eq. (\ref{eq:fidel}).
%As defined earlier, it is clear that 
However, finding optimal fidelity after maximizing over measurements and unitaries in a multipartite scenario is not easy. As a way-out, looking at the symmetries of the states involved, we choose a particular kind of measurement and unitary operators,  which provides a  lower bound on ${\cal F}(\rho)$. 
In this paper, we deal with  two information-theoretically important families of three-qubit pure states, 
%. We find that TD-TC protocol reveals a new classification of three-qubit pure states according to their usefulness in this scheme. For such study we choose the shared state 
% among Alice and Claires and among Bob and Claires are either the family of  %,  of entangled three qubit pure states. 
%In this paper we analyse the performance of the channels with a single block. As for the 
%distribution and concentration units in the block, we choose from the three party pure 
%SLOCC classes of entangled states. 
the generalized Greenberger-Horne-Zeillinger (GHZ)  \cite{gGHZ}, and  the generalized W states \cite{dur2000}, shared between $(A, C_1, C_2)$ and \((C_1, C_2, B)\)-trios. 
%, belonging to the two distinct SLOCC 
%classes \cite{dur2000} of three qubit pure entangled states.
In the subsequent calculations, we do not explicitly mention about the time delay issue but only consider the fidelity yield of various configurations considered in this manuscript. However, note that all these strategies can be easily continued to fit the conditions of delayed teleportation.

%The arbitrary state, $|\psi\rangle$, that Alice wants to send is  given by
%\begin{equation}\label{eq:input}
% \ket{\psi}=a\ket{0}+b\ket{1},
%\end{equation}
%where $|a|^2 + |b|^2 = 1$.
%, to Bob. Now we will consider the lower bound of ${\cal F}(\rho)$,
%for two different choices of three qubit quantum states.

%\begin{figure}
%\begin{center}
% \includegraphics[scale=0.8]{gGHZc1.pdf}
% \caption{ Schematic diagram of quantum teleportation using a three party quantum channel
% state.}\label{fig:gGHZ}
%\end{center} 
% \end{figure}

 \section{Two-path quantum repeater:  Noiseless scenario}
 \label{sec:dis_conc_perform}
%%%%%%%%%%%%%%%%%%%%%%%%%%%%%%%%%%%%%%%%%%%%%%%%%%%%%%%%%%%%%%%%%%%%%%%%%%%%%%%%%%%%
In this section, we assume that the shared three-qubit states used as quantum channels are noiseless, and we explicitly discuss the protocols as well as the evaluation of their fidelities. The effects of local noise on the  protocol will be discussed in the succeeding sections.

\subsection{TD-TC protocol via Generalized GHZ state}
\label{gGHZ}

%\subsubsection{Distribution and concentration}

Let us first consider the situation, in which Alice shares a three-qubit generalized GHZ state 
$\ket{gGHZ}$, given by
\begin{equation}
 \ket{gGHZ (\alpha)} =\sqrt{\alpha}\ket{000}+\sqrt{1-\alpha} ~e^{i\phi}\ket{111},
 \label{dcchannels}
\end{equation}
where $\alpha \in (0,1)$ and $\phi \in [0,2\pi)$,  with  $C_1$ and $C_2$. The receiver,  Bob, shares  another copy of the same state, $\ket{gGHZ (\alpha)}$, with the auxiliary nodes, $C_1$ and $C_2$. 
% as shown in figure \ref{fig:gGHZ}. 
%The three-qubit gGHZ state reads as
Note here that all the parties except  Bob 
(to whom the unknown state has to be teleported) initially possess two qubits.
%, Alice posses one unknown input state
%$|\psi\rangle$, and another from the  $\ket{gGHZ_1}$. 
%Similarly,  both the Claires possess two subsystems, one from $\ket{gGHZ_1}$
%and another one from $\ket{gGHZ_2}$. 
%We also assume that
%\begin{equation}
% \ket{gGHZ_1}=\ket{gGHZ_2}=\ket{gGHZ}_{A(B)C_1C_2}.
%\end{equation}
%Now we use a particular kind of measurement and a fix choice of unitary rotations, which is stated below.
Alice first performs a joint Bell measurement $\{|\phi^\pm\rangle, |\psi^\pm\rangle\}$
\cite{Bell-basis}  on the input state \(\ket{\psi}\), and the subsystem of $|gGHZ (\alpha)\rangle$ in her possession. The unnormalized post measured states (PMS) at \(C_1\) and \(C_2\)
% $|\xi_{\phi}^\pm\rangle_{C_1C_2}$
%and $|\xi_{\psi}^\pm\rangle_{C_1C_2}$, where 
read as
\begin{eqnarray}
|\xi_{\phi}^\pm\rangle_{C_1C_2} &=&  {}_{A'A}\langle \phi^\pm| (\ket{\psi}_{A'} \otimes \ket{gGHZ}_{AC_1C_2}), \nonumber \\
&=& (a\sqrt{\alpha}|00\rangle \pm b \sqrt{1-\alpha}e^{i\phi}|11\rangle)/\sqrt{2} \label{eq:PMS1}\\
|\xi_{\psi}^\pm\rangle_{C_1C_2} &=&  {}_{A'A}\langle \psi^\pm| (\ket{\psi}_{A'} \otimes \ket{gGHZ}_{AC_1C_2}), \nonumber \\
&=& (b\sqrt{\alpha}|00\rangle \pm a \sqrt{1-\alpha}e^{i\phi}|11\rangle) /\sqrt{2}.
\label{eq:PMS2}
\end{eqnarray}
%Next, she communicates her measurement outcomes to both the Claires over classical channel, and
 Depending on the measurement outcomes obtained and communicated by Alice,  the Claires
perform local unitary operations, chosen from the Pauli operators,  $\{I, \sigma_x, \sigma_y,
\sigma_z\}$ on their respective parts of the shared $|gGHZ (\alpha)\rangle$ state. 
%The  unitaries are suitably chosen from the set of Pauli matrices $\{I, \sigma_x, \sigma_y,
%\sigma_z\}$\footnote{The three Pauli matrices are given by
%$$
% \sigma_x=\left(
%\begin{array}{cc}
%0 & 1 \\
%1 & 0
%\end{array}\right)
%,~~~~~~~~~~
%\sigma_y=\left(
%\begin{array}{cc}
%0 & -i \\
%i & 0
%\end{array}\right)
%,~~~~~~~~~~
%\sigma_z=\left(
%\begin{array}{cc}
%1 & 0 \\
%0 & -1
%\end{array}\right)
%.$$
%}, and t
The set of local unitaries at the end of \(C_1\) and \(C_2\)  can jointly be represented as  $\{I \otimes I, I\otimes \sigma_z, \sigma_x\otimes \sigma_x, \sigma_x\otimes \sigma_y\}$, corresponding to  $\{|\phi^\pm\rangle, |\psi^\pm\rangle\}$ clicks.
%listed in the Table \ref{T:uniraty_ghz}.
%%The choices of unitary operators are given in the Table below.
%\begin{table}[ph]
%\begin{center}
%\begin{tabular}{| C{1.7cm} | C{1.7cm} | C{1.7cm} | N}
%\hline
%\cellcolor{LightRed}{\bf Outcome} & \multicolumn{2}{|c|}{\cellcolor{Yellow}{ \bf Unitary operators}} & \\ [1.2ex]
%\hline
% Alice  & Claire 1 & Claire 2 &\\ [1.2ex]
%\hline
% $|\phi^+\rangle$ & $I$ & $I$ &\\ [1.2ex]
%\hline
%$|\phi^-\rangle$ & $I$ & $\sigma_z$ &\\ [1.2ex]
%\hline
%$|\psi^+\rangle$ & $\sigma_x$ & $\sigma_x$ &\\ [1.2ex]
%\hline
%$|\psi^-\rangle$ & $\sigma_x$ & $\sigma_y$ &\\ [1.2ex]
%\hline
%\end{tabular}
%\caption{Table of the unitary rotations by Clairs corresponding to the possible  outcomes
%of the  Bell basis measurement by Alice, when they share a multiparty gGHZ state.}
%\label{T:uniraty_ghz}
%\end{center}
%\end{table}

%In this step, we first departed from the standard protocol of teleportation as given in \cite{bennett1993}.
After performing these unitary operations, the PMS shared by $C_1$ and $C_2$ are the 
$|\xi_{\phi}^+\rangle_{C_1C_2}$ and $|\xi_{\psi}^+\rangle_{C_1C_2}$, without normalization.
% It is worth noting that,
%the PMS's are not a physical quantum state, as we have not normalized them. 
Note that the normalization constants of these states are the probabilities of obtaining the outcomes of the Bell measurement performed by Alice.
The first two steps is a part of TD protocol. After this,  $C_1$ and $C_2$ perform two independent Bell measurements on their auxiliary nodes, communicate their measurement results to Bob and finally Bob chooses unitary operators given in Ref. \cite{table_ghz}, depending on the eight measurement outcomes $\{|\phi^\pm\rangle \otimes |\phi^\pm\rangle\}$ and $\{|\psi^\pm\rangle \otimes |\psi^\pm\rangle\}$.
%So we can consider that the states in Eqs. (\ref{eq:PMS1}) and (\ref{eq:PMS2}) are holding the information of the probability of occurrence within them.
 %Now both the Claires share another generalized GHZ state with Bob. 
%In other words, each Claire is in possession of two subsystems, belonging to two different three 
%party entangled states. Claire 1 now performs a Bell measurement on the two subsystems she has
%and also Claire 2 performs another Bell measurement on her subsystems. 
%%The choice of the measurement by $C_2$ is completely  independent of the measurement done by Claire 1.
%They then individually communicate the measurement outcome to Bob, and Bob performs unitary rotation on his part of the shared state to obtain the  final state.
 An interesting point to note here is that the structure of the gGHZ  state guarantees that if  $\ket{\phi^\pm}$ clicks at \(C_1\)'s port, then 
$\ket{\psi^\pm}$ can never click at \(C_2\)'s end and vice-versa. Thus we can have only eight possible
outcomes of the measurements instead of sixteen, which would later turn out to be useful in a noisy scenario. 
In each of the cases, the (unnormalized) quantum state, teleported to Bob, is given in  Table \ref{T:outcome_ghz}.

\begin{table}[h]
\begin{center}
\begin{tabular}{| C{1.3cm} | C{1.3cm} | C{4.7cm} | N}
\hline
 \multicolumn{2}{|c|}{\cellcolor{LightRed}{\bf Outcomes by}} & \cellcolor{LightGreen}{\bf Teleported state to} &\\ [1.2ex]
\hline
  \rowcolor{Yellow}{$C_1$} & $C_2$ & Bob ($B$) &\\ [1.2ex]
\hline
\rowcolor{LightCyan}
 \multicolumn{3}{|c|}{When Alice obtains $|\phi^\pm\rangle$} &\\ [1.2ex]
 \hline
 $|\phi^\pm\rangle$ & $|\phi^\pm\rangle$ & $(a\alpha |0\rangle + b(1-\alpha)e^{2i\phi}|1\rangle)/2\sqrt{2}$ &\\ [1.2ex]
\hline
$|\psi^\pm\rangle$ & $|\psi^\pm\rangle$ & $\sqrt{\alpha (1-\alpha)}e^{i\phi}(a |0\rangle + b|1\rangle)/2\sqrt{2}$ &\\ [1.2ex]
\hline
\rowcolor{LightCyan}
\multicolumn{3}{|c|}{When Alice obtains $|\psi^\pm\rangle$} &\\ [1.2ex]
\hline
$|\phi^\pm\rangle$ & $|\phi^\pm\rangle$ & $\sqrt{\alpha (1-\alpha)}e^{i\phi}(a |0\rangle + b|1\rangle)/2\sqrt{2}$ &\\ [1.2ex]
\hline
$|\psi^\pm\rangle$ & $|\psi^\pm\rangle$ & $(a(1-\alpha)e^{2i\phi} |0\rangle + b\alpha|1\rangle)/2\sqrt{2}$ &\\ [1.2ex]
\hline
\end{tabular}
\caption{Table of  four possible teleported states at the output port, B, after different measurement
outcomes by Alice and Claires, $C_1$ and $C_2$ and the unitary rotations by Claires and Bob.}
\label{T:outcome_ghz}
\end{center}
\end{table}
Therefore, the fidelity in multipath QR by using gGHZ state  is constrained by
\begin{eqnarray}
{\cal F}^{DC}(gGHZ) &\geq & \int d^2a d^2b \Big( \big| |a|^2\alpha + |b|^2(1-\alpha)e^{2i\phi} \big|^2 \nonumber \\
&& + 2 \alpha(1-\alpha)  + \big| |a|^2(1-\alpha)e^{2i\phi} + |b|^2\alpha \big|^2 \Big), \nonumber \\
&=& \frac 23 + \frac 43 \alpha(1-\alpha)\cos^2\phi.
\label{fid_withphase}
\end{eqnarray}
%Note that the second term of the above equation is always 
%positive except when $\phi = \frac{\pi}{2}$ and hence we conclude that it is always possible to faithfully teleport an unknown quantum state to Bob  in the TD and TC scheme with the shared two copies of gGHZ state in a better way than any classical protocol. Now $F \leq {\cal F}({gGHZ})$, defined in Eq. (\ref{eq:fidel}). This lower bound 
The lower bound on  ${\cal F}^{DC}({gGHZ})$ can be improved if one  absorbs
the phase factor of the gGHZ state to any one of the measurements by the Claires or in 
any one of the unitary operations by Claires or Bob. For example, in the Bell  measurement, performed 
by Alice,  she  can choose the basis by redefining $|1\rangle \rightarrow \tilde{|1\rangle}=  e^{i\phi}|1\rangle$. Immediately,  Eq.  (\ref{fid_withphase}) reduces to
\begin{equation}\label{eq:fidel_gGHZ_TDTC}
{\cal F}^{DC}(gGHZ (\alpha)) \geq \frac 23 + \frac 43 \alpha(1-\alpha) \equiv F(gGHZ).
\end{equation}
%The quantity in the right hand side provides the lower bound of ${\cal F}_{gGHZ}$. I
Notice that it reaches to unity, when $\alpha = \frac 12$, implying the protocol to be optimal when the shared multiparty quantum state is the GHZ state. It is tempting to conjecture at this point that the TD-TC protocol described above is possibly the optimal one also for the gGHZ state. Furthermore, we would show in a later section that the fidelity obtained above is independent of the the number of intermediate Claires.

\begin{figure}
\begin{center}
\includegraphics[scale=0.75]{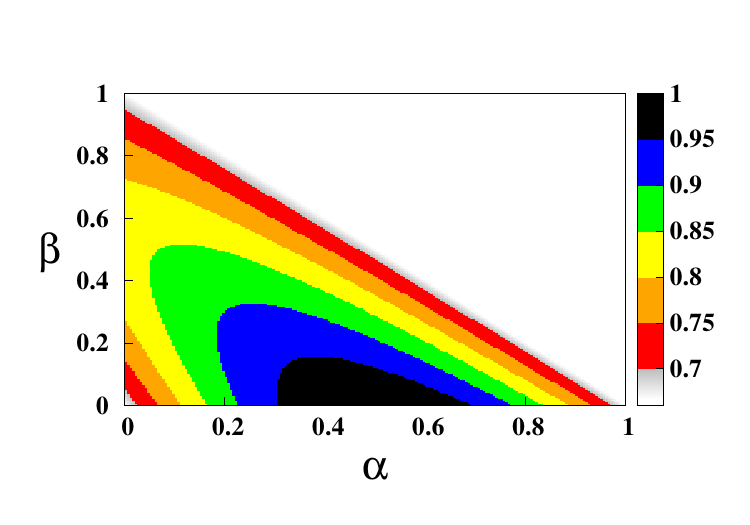}
\end{center}
  \caption{Map of  fidelity of a generalized W state vs. state parameters, $\alpha$ and $\beta$, in the multipath QR process. 
 %for a particular choice of meassurement  settings and unitary rotations. 
 Both axes are dimensionless.}\label{fid_plot_W}
\end{figure}

%%%%%%%%%%%%%%%%%%%%%%%%%%%%%%%%%%%%%%%%%%%%%%%%%%%%%%%%%%%%%%%%%%%%%%%%%%%%%%%%%%%%%%
\subsection{Generalized W states as multipath quantum channels}\label{W}
%%%%%%%%%%%%%%%%%%%%%%%%%%%%%%%%%%%%%%%%%%%%%%%%%%%%%%%%%%%%%%%%%%%%%%%%%%%%%%%%%%%%%%

%\subsubsection{Distributiona and concentraion}

We now move on to a scenario, where the distribution and concentration channels are from the two-parameter family of  three-qubit
generalized W states \cite{dur2000}, given by 
\begin{equation}\label{Eq:gW}
 \ket{gW (\alpha, \beta)}=\sqrt{\alpha}\ket{001}+\sqrt{\beta}\ket{010}+\sqrt{1-\alpha-\beta}\ket{100},
\end{equation}
where $\alpha, \beta \in (0,1)$ and $\alpha + \beta < 1$.
Like the case of the gGHZ state, \((A, C_1, C_2)\) and \((C_1, C_2, B)\) share two copies of gW state, 
%here we also assume that  
$\ket{gW (\alpha, \beta)}$. 
%, and $|gW_2\rangle$ are shared
%between Alice and  Claires as well as the Claires and Bob respectively.
%. As in the case of the Teleporatation with generalized GHZ states, we consider that the states used as the distribution and concentration  
%\begin{equation}
%\ket{gW_1}=\ket{gW_2}=\ket{gW}_{A(B)C_1C_2}.
%\end{equation}
At the input port and at both  the auxiliary nodes,  Bell  measurements are carried out as before. However, the choice of unitary operators are different than that in the preceding section.
% and the input state is $|\psi\rangle$ given
%in Eq. (\ref{eq:input}).
%After Alice initiates the measurement on the joint system in her possession, the unnormalized post
%measurement states can be represented as
%Alice communicates her measurement outcomes to the Claires, who 
\(C_1\) and \(C_2\) perform identities if $\ket{\phi^+}$ or $\ket{\psi^+}$ clicks at Alice's node while for the rest of the measurement outcomes,  they operate $\sigma_z$ in their subsystems.
%. The choices of unitaries of the Claires
%are given below.
%\begin{table}[ht]
%\begin{center}
%\begin{tabular}{| C{1.7cm} | C{1.7cm} | C{1.7cm} | N}
%\hline
%\cellcolor{LightRed}{\bf Outcome} & \multicolumn{2}{|c|}{\cellcolor{Yellow}{\bf Unitary operators}} &\\[1ex]
%\hline
% Alice  & Clair 1 & Clair 2 &\\ [1ex]
%\hline
% $|\phi^+\rangle$ & $I$ & $I$ &\\ [1ex]
%\hline
%$|\phi^-\rangle$ & $\sigma_z$ & $\sigma_z$ &\\ [1ex]
%\hline
%$|\psi^+\rangle$ & $I$ & $I$ &\\ [1ex]
%\hline
%$|\psi^-\rangle$ & $\sigma_z$ & $\sigma_z$ &\\ [1ex]
%\hline
%\end{tabular}
%\caption{Table of the unitary rotations by Claires corresponding to the possible  outcomes of the  Bell 
%basis measurement by Alice, when they share a multiparty gW state.}
%\label{T:uniraty_gW}
%\end{center}
%\end{table} 
The applications of these local unitaries reduce the PMS from four to two, shared by $C_1$ and $C_2$, and are given by
% they are
\begin{eqnarray}
&&  \hspace{-1.1em} |\zeta_{\phi}^{+}\rangle_{C_1C_2} 
% {}_{A'A}\langle \phi^\pm| (\ket{\psi}_{A'} \otimes \ket{gW}_{AC_1C_2}), \nonumber \\
  = \frac{1}{\sqrt{2} }\Big(a \big(\sqrt{\alpha}|01\rangle +
 \sqrt{\beta}|10\rangle \big) + b\sqrt{1 - \alpha - \beta}|00\rangle \Big), \nonumber \\
 ~~\label{eq:PMSW1}\\
&& \hspace{-1.1em} |\zeta_{\psi}^{+}\rangle_{C_1C_2} =  
%{}_{A'A}\langle \psi^\pm| (\ket{\psi}_{A'} \otimes \ket{gW}_{AC_1C_2}), \nonumber \\
 \frac{1}{\sqrt{2}}\Big( a\sqrt{1 - \alpha - \beta}|00\rangle  +
 b \big(\sqrt{\alpha}|01\rangle + \sqrt{\beta}|10\rangle \big)  \Big).\nonumber \\~~
\label{eq:PMSW2}
\end{eqnarray} 
%$|\zeta_{\phi}^+\rangle_{C_1C_2}$ and $|\zeta_{\psi}^+\rangle_{C_1C_2}$. 
%After TD part, each of the Claires now performs a Bell measurement on their subsystems, followed by the
In  the TC part,  local unitaries on the output port, $B$, are given in Table \ref{Table:secondmeasure} of Appendix \ref{sec:outcome_gW_dis_conc}, depending on the results  of two Bell measurements, executed by \(C_1\) and \(C_2\).
%belonging to the 
%distribution and the concentration 
%channels and then communicates her outcomes classiclally to Bob. Upon receiving the result
%Bob applies local unitaries on the output port go get the final state. 
%Owing to two different joint states with the Claires at the end to the distribution phase, the concentration 
%protocol is much more complicated and is best summed up using table~\ref{Table:secondmeasure} in Appendix \ref{sec:outcome_gW_dis_conc}.
Therefore, the fidelity of the TD-TC protocol, described above, for the shared generalized W states can then be estimated as 
\begin{eqnarray}
{\cal F}^{DC}(gW (\alpha, \beta)) \geq  F(gW) &=& 2  \int d^2 a d^2b  \Bigg[ \alpha \beta (|a|^4 + |b|^4 ) + \frac{1}{4} \bigg(\Big(|a |^2 (\alpha + \beta) +
|b |^2 (1-  \alpha - \beta) \Big)^2 \\\nonumber
&& + \Big(|a |^2 (\alpha - \beta) +
|b |^2 (1-  \alpha - \beta) \Big)^2 + \Big(|a |^2 (1-  \alpha - \beta) + |b |^2 (\alpha + \beta) \Big)^2 \nonumber \\    
&&+ \Big(|a |^2 (1-  \alpha - \beta) + |b |^2 (\alpha - \beta) \Big)^2 \bigg) \Bigg] + 2 (\alpha + \beta) (1 - \alpha - \beta), \nonumber \\
&=& \frac{2}{3} + \frac{2}{3}( 2 \alpha + \beta)(1 - \alpha -\beta). \label{eq:fidel_gW_TDTC}
\end{eqnarray}

%The probabilities of all 
%the possible outcomes are calculated and used in obtaining the 
%actual fidelity of the output state for the input state $\ket{\psi}$. Integrating over 
%all the possible input states, the average fidelity for the protocol using generalized W
%state as distribution and concentration channels is obtained. The average fidelity for the
%channel is given by
%\begin{equation}
% F_{gW}=\frac{2}{3} + \frac{2}{3}\Big( 2 \alpha(1 - \alpha) + \beta(1 - \beta) - 3\alpha\beta \Big)
% \label{fid_W}
%\end{equation}
%For the detailed calculation of the the fidelity, please see Appendix.
\emph{Remark.} The behaviour of fidelity for the generalized W state in the $(\alpha, \beta)$-plane is depicted in Fig. \ref{fid_plot_W}. 
From the figure, it is clear that the  choice of measurements and unitaries discussed above  leads to a maximal  fidelity in a region where $\alpha \approx \frac 12$ and $\beta \approx 0$. This situation arises when the shared gW state
% has almost vanishing genuine multiparty entanglement due to the fact that it 
is close to a product state in the  $A(B)C_2:C_1$ bipartitions, having negligible genuine multipartite entanglement. 
%One can show that if the state is almost product in the 
A similar scenario is true when the state is product across $A(B)C_1:C_2$ bipartition, although the choice of unitary operators in this case needs to be different.
% the fidelity almost reaches its maximum value with different choices of unitary operators corresponding to the outcomes of Bell measurements.

%In the next section, we compare this distribution and concentration protocol with the case when every
%multiparty repeaters block converted into a linear chain repeater by either ignoring the presence of 
%other parties or performing individual optimal measurements in the other ports,  as shown in figure~\ref{gGHZ_m}. Moreover, we will 
%also analyze the presence of different kind of multiparty channel and it is interesting to notice that 
%the choice of channel can differentiate the two kind of linear chains constructed the from above stated
%two methods.

\begin{figure}
\begin{center}
\includegraphics[scale=0.8]{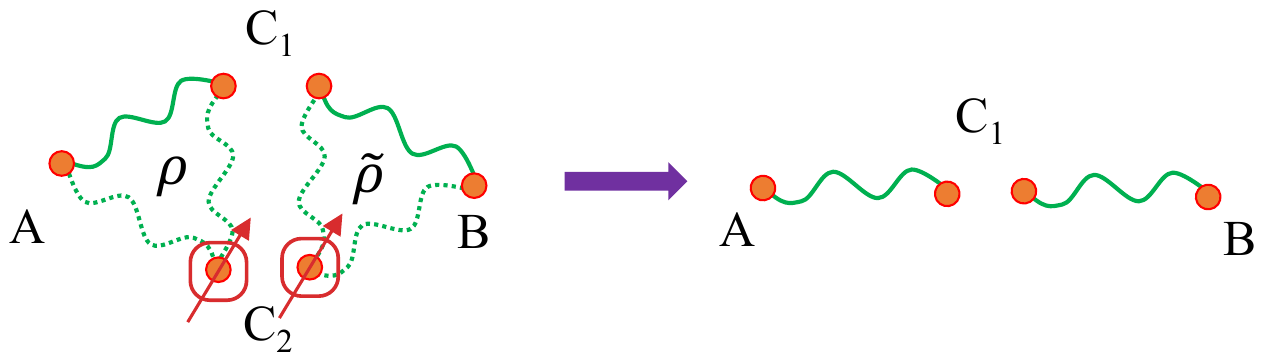}
 \end{center}
 \caption{\label{fig:measurementsingle} 
 Local measurement-based single-path QR. Optimal local measurements are performed by \(C_2\), reducing  tripartite channels to  bipartite ones. For example, executing local measurements by \(C_2\) disentangles \(C_2\) from \(A\) and \(C_1\) and the entire TD-TC scheme reduces  to the one consisting of \((A,C_1)\) and \((C_1, B)\)  duos. 
 %A simple case of converting teldistribution and teleconcentration
% unit into a linear chain with one intermediate party. This is achieved if one of the 
% intermediate parties, performs optimal measurements, in this case in the 
% $\{\frac{\ket{0}+\ket{1}}{\sqrt{2}}, \frac{\ket{0}-\ket{1}}{\sqrt{2}}\}$ basis, on her 
% subsystems, as shown in the figure.
}
 \end{figure}

%Instead of TD and TC scenario, we now consider a scheme where one of the Claire initially performs optimal local measurements in the shared state, thereby reducing the protocol to a linear chain of entanglement swapping. Our aim is to compare TD-TC protocol with the linear chain protocol.

\section{Local measurement-based Single-Path quantum repeater}
\label{sec:linearchain}
Instead of the TD-TC scenario, we now consider a scheme where one of the Claires initially performs optimal local measurements in the shared state, thereby reducing the protocol consisting of  two linear quantum channels.
As discussed earlier, such a situation is important if one of the Claires decline active participation in the TD-TC scheme and leaves the protocol before its implementation. Note that this would decrease the guarantee of the perfect timing of the protocol completion, since there is only one Claire left.  Nevertheless, 
 our aim here is to compare fidelities obtained by multipath  protocol with that of the local measurement-based single-path  ones.
% we will modify the TD-TC protocol in such a way that multiparty protocol reduces to a bipartite repeater protocol. 
Specifically, 
%consider a slightly different variation of the teleportation protocol involving 
%quantum repeaters \cite{q_repeater}. In the previous section, we looked at the scenario where 
%teleportaion is carried out by using two multipartite states, one shared between the sender and the 
%intermediaries and the other shared between the receiver and the latter. In this case, however, we
%replace the multiple intermediaries by a single party, say Claire 1. The state shared between the 
%sender, Alice and Claire 1 as well as that shared between Claire 1 and the receiver, Bob, is derived
from three-party quantum states, $\rho$ and $\tilde{\rho}$, shared between \((A, C_1, C_2)\) and \((C_1, C_2, B)\) trios,  certain two-party states, $\sigma_{\rho}$ and $\tilde{\sigma}_{\tilde{\rho}}$ are obtained by operating suitable rank-one projective measurements by one of the Claires, say, \(C_2\) (see Fig. \ref{fig:measurementsingle}). In this case, \(C_1\) which actively
participates in the protocol can be called functioning port where as $C_2$ is the non-functioning port. As in the previous case, initially, either  two copies of the
generalized GHZ or the generalized W state as  multiparty channels are shared among all the parties. 

The motivation behind considering such a protocol is as follows. First of all,  there can be a situation where one of the parties,  say $C_2$, for some reason, has to leave the protocol, reducing the initial pure state to a mixed state. In case of the gGHZ state, such scenario leads to a separable state between $A$ and $C_1$ as well as $C_1$ and $B$, although for the gW state, the reduced state, $P[\sqrt{\beta}~|01\rangle + \sqrt{1- \alpha - \beta}~|10\rangle] + \alpha P[|00\rangle]$, where $P[\ket{\psi}] = \ket{\psi}\bra{\psi}$ is entangled and hence can be used for teleportation protocol with quantum fidelity. The example of the gGHZ state shows that such betrayal or absence can sometimes  eliminate the quantum advantage from the communication protocol.
Under such circumstances, a profitable situation can be regained if one considers a local measurement-based protocol, if the situation is not of betrayal and the absent party can pre-measure in some basis of his/her parts of the system.
In particular, instead of leaving the protocol, $C_2$ can help to share highly entangled state to $(A,C_1)$ and $(C_1,B)$ pairs by performing suitable local measurements on her parts,  transforming tripartite states to bipartite ones, which are useful for quantum teleportation.
% Then the protocol then turns to be more beneficial and we refer this scheme as measurement-based single-path quantum repeater, consisting of several linear chains.
%In this case, $C_2$ can be called partially functioning port. We now compare the original TD-TC protocol with the single path quantum repeater where fidelity is obtained after $C_2$ performs optimal local measurements and we will show that multiparty protocol sometimes behaves better than that of the linear chain. 
In the local measurement-based scenario, after optimization performed over local measurements, the optimal fidelity can be obtained by using the formula of teleportation fidelity in terms of singlet fraction \cite{horo_fid}. Note that, similar motivation leads to the concept of entanglement assistance or localizable entanglement \cite{baby}.

\begin{table}
\begin{tabular}{|M{1.5cm}|M{1.75cm}|M{8cm}|M{3.8cm}|N}
\hline
{\bf Outcome} & {\bf Probability} & {\bf Normalised PMS} & {\bf Schmidt form} &\\ [1.5ex]
\hline 
$|M\rangle$ & $p_M $ &  $\frac{1}{\sqrt{p_M}}(e^{-i\phi}\sqrt{\alpha(1-x)}|00\rangle +\sqrt{\beta x}|01\rangle +\sqrt{x(1-\alpha-\beta)}|10\rangle)$ & $\sqrt{\Lambda^+_M}|\tilde{0}\bar{0}\rangle + \sqrt{\Lambda^-_M}|\tilde{1}\bar{1}\rangle$ & \\ [4.3ex]
\hline  
 $|M^\perp\rangle$ & $p_M^\perp $ & $\frac{1}{\sqrt{p_M^\perp}}\sqrt{\alpha x}|00\rangle +e^{i\phi}\sqrt{\beta (1-x)}|01\rangle +\sqrt{(1-x)(1-\alpha-\beta)}|10\rangle$ & $\sqrt{\Lambda^+_{M^\perp}}|\tilde{0}\bar{0}\rangle + \sqrt{\Lambda^-_{M^\perp}}|\tilde{1}\bar{1}\rangle$ & \\ [4.3ex]
\hline 
\end{tabular}
\caption{Table of the post measured states (PMS) and the corresponding probabilities, 
 $p_M =\alpha +x -2\alpha x$ and $p_M^\perp = 1-(\alpha +x -2\alpha x)$. 
$\Lambda^{\pm}_{M}$ and $\Lambda^{\pm}_{M^\perp}$ are the Schmidt numbers 
\cite{Schmidt_decomposition} for PMS when $M$ and $M^\perp$ clicks, and are given by 
$\Lambda^{\pm}_{M} = \frac{1}{2}\left(1 \pm \sqrt{1 - \frac{4x^2\beta(1-\alpha-\beta)}
{(\alpha +x -2\alpha x)^2}}\right)$ and $\Lambda^{\pm}_{M^\perp} 
= \frac{1}{2}\left(1 \pm \sqrt{1 - \frac{4(1-x)^2\beta(1-\alpha-\beta)}{(1-(\alpha +x -2\alpha x))^2}}\right)$.
The corresponding Schmidt vectors are labeled as 
$\lbrace\ket{\tilde{0}},\ket{\tilde{1}}\rbrace$ and $\lbrace\ket{\bar{0}},\ket{\bar{1}}\rbrace$
for Alice and $C_1$ respectively.}
\label{t:gw_opt}
\end{table}

\subsection{Generalized GHZ  vs. generalized W states for local measurement-based scheme}\label{sec:dis_con_gGHZ}

Let us start the discussion with a scenario where  an arbitrary three-qubit  gGHZ state is shared between the Alice (Bob) and Claires. 
%then we notice the following. \\
%{\bf Theorem1:} \textit{For a shared gGHZ state, performing measurement on the non-functioning port
%is always better than ignorance.}\\
%will now show that performing local measurement by rest of the Claires is better than ignorance. 
As mentioned,  complete ignorance in the non-functioning port leads to a two party reduced 
density matrix,  given by
\begin{equation}
\rho_{A(B)C_1}^{gGHZ} = \alpha |00\rangle\langle 00| + (1 - \alpha)|11\rangle \langle11|.
\end{equation}
As it is a separable state in the $A(B):C_1$ bipartition,
% the  fidelity for the above channel  is 
${\cal F}^l_1 = \frac 23$. 
On the other hand, if  $C_2$ chooses an arbitrary rank-one projective measurement, given by
\begin{eqnarray}
|M\rangle = \sqrt{x}|0\rangle + e^{i\theta}\sqrt{1-x}|1\rangle, \label{eq:m_basis1} \\
 |M^\perp\rangle = \sqrt{1-x}|0\rangle - \sqrt{x}e^{i\theta}|1\rangle, \label{eq:m_basis2}
\end{eqnarray}
the PMS between Alice (Bob) and \(C_1\) can be represented as 
\begin{eqnarray}
|\xi_M\rangle = \frac{1}{\sqrt{p_M}}(\sqrt{x\alpha}|00\rangle + e^{i(\phi -\theta)}\sqrt{(1-x)(1 - \alpha)}|11\rangle), ~~~~~\\ 
|\xi_M^\perp\rangle = \frac{1}{\sqrt{p_M^\perp}}(\sqrt{x(1-\alpha)}|00\rangle - e^{i(\phi -\theta)}\sqrt{x(1 - \alpha)}|11\rangle),~~~~~
\label{eq:m_basis_outcome}
\end{eqnarray}
where $p_M = x\alpha + (1-x)(1 - \alpha)$ is the probability of obtaining $|M\rangle$ and $p_M^\perp = x(1-\alpha) + (1-x)\alpha$ for $|M^\perp\rangle$. 
 The  teleportation  fidelity by using PMS as channels between Alice to $C_1$ is \cite{horo_fid}
\begin{eqnarray}
f_{max} = \frac 23 + \frac 23 \sqrt{x \alpha (1-x)(1-\alpha)},
\label{eq:fmax}
\end{eqnarray}
which reaches its maximum value for $x = \frac 12$, representing the optimal measurement basis as $\ket{\pm} = \frac{1}{\sqrt{2}}(\ket{0} \pm \ket{1})$.
We now assume that the teleportation fidelity, by using single-path QR, attains its maximum value when each
segment can teleport at its maximum capacity.  
%i.e, when the segmented channel has the highest teleportation fidelity. Moreover, the joint measurements and the corresponding unitary rotations which leads to that maximum teleportation fidelity, here for two repeater scenario we use the same measurements and unitary for both the channels.
Starting from the  gGHZ state, optimal measurements on both parts of the shared state by $C_2$ lead to the bipartite states shared between Alice (Bob) and $C_1$, given by 
\begin{equation}
|\mu^\pm\rangle_{A(B)C_1}^{gGHZ} = \sqrt{\alpha}|00\rangle \pm \sqrt{1-\alpha} e^{i\phi}|11\rangle,
\end{equation}
with probability $\frac 12$. 
%, and they are connected by a local unitaries.
 The fidelity of single-path QR, via quantum channels, $(A,C_1)$ and $(C_1,B)$ pairs, finally reads 
\begin{equation}\label{eq:fidel_gGHZ_lin}
{\cal F}^l_2(gGHZ) = \frac 23 + \frac 43 \alpha(1 - \alpha).
\end{equation}
For the details, see \ref{sec:two_linearchain}.

%If we compare the fidelity of the single path repeater, obtained after suitable measurement, with the fidelity of TD and TC protocol i.e., 
Comparing Eqs. (\ref{eq:fidel_gGHZ_TDTC}) and (\ref{eq:fidel_gGHZ_lin}), 
we find that the fidelities obtained by the specific TD-TC protocol with that of the local measurement-based ones exactly match and therefore,  we have 
\begin{equation}
{\cal F}^{DC}(gGHZ) \geq {\cal F}_{2}^{l}(gGHZ) = F (gGHZ),
\end{equation}
where the equality holds for the GHZ state with $\alpha = \frac 12$. 
%One should note here that the  inequality remains for the entire family of the gGHZ state execpt the GHZ point,  since the fidelity obtained in the TD-TC protocol may not be optimal.
 If we now assume that with the shared gGHZ state, the protocol presented in Sec \ref{sec:dis_conc_perform} is the optimal one,  we have that multipath QR in terms of teleportation fidelity does not provide any advantage  over local measurement-based single-path QR.
We will show that when the shared state is the gW state,  two scenarios no more remains the  same.
%which we will consider in the next section.
% in pratical situations, in the presence of noise, a problem we take up later.

%\subsection{Measurement starts from Generalized W state}
%Let us now consider the situation of measurement-based single-path quantum repeater  when
 Let us now move to the protocol with the initial state being the generalized W state.   Let us first state the following proposition. \\
{\bf Proposition:} {\it Consider two copies of a generalized W state shared between $A,C_1,C_2$ and $C_1,C_2,B$. Suppose that $C_2$ must remain non-functional during the actual implementation of the teleportation, but agrees to make measurements before. The optimal fidelity in the single-path quantum repeater can be obtained when the local measurements at $C_2$ are performed in the computational basis.}\\
\textsf{Proof:} 
%If Claire 2 performs measurements in any arbitrary basis, 
%on her individual parts of the shared states, then the optimized measurements are the computational 
%basis $\{|0\rangle, |1\rangle\}$ measurements. To prove the theorem, we use the fact that the fidelity 
%of teleportation, in the linear chain, is maximized when in each segment one can have the best possible
%bipartite quantum state. So here we just prove that by performing measurement, on any one of the shared
%state, in the computational basis, Claire 2 can help Alice and Claire 1, to share the best possible quantum
%state. This result also follows for Bob and Claire 1.
After performing  measurements by $C_2$ in an arbitrary basis given in Eqs. (\ref{eq:m_basis1}) and (\ref{eq:m_basis2}),  the PMS and their 
probabilities of occurrence can be computed (see Table \ref{t:gw_opt}).
The maximum teleportation fidelity, $f_{max}^{gW}$, via PMS from A to $C_1$  in terms of the maximal singlet 
fraction \cite{horo_fid}, $F_{max}$,  is given by
\begin{eqnarray}
f_{max}^{gW} = p_Mf^M_{max} + p_M^\perp f^{M^\perp}_{max},
\label{eq:fmax_def}
\end{eqnarray}
where  
\begin{eqnarray}
f_{max}^{M(M^\perp)} &=& \frac{2F_{max}^{M(M^\perp)}+1}{3}, ~~ \text{and} \label{eq:fidel_singletfrac}  \\
F_{max}^{M(M^\perp)} &=& \frac{1}{2}\Big( 1 + 2\sqrt{\Lambda^+_{M(M^\perp)}\Lambda^-_{M(M^\perp)}} \Big)\label{eq:singlet_frac}.
\end{eqnarray}
 Substituting values from Table. \ref{t:gw_opt}, and maximizing over the measurement basis,  we obtain
\begin{eqnarray}
f_{max}^{gW} = \frac{2}{3}+\frac{2}{3}\sqrt{\beta(1-\alpha-\beta)}.
\end{eqnarray}
%The optimal  measurement basis, chosen by 
We now note that if  \(C_2\) makes the measurement in the $\{\ket{0}, \ket{1}\}$ basis, the PMS reduces to
\begin{eqnarray}
&&|\zeta_0\rangle = \frac{1}{\sqrt{p_0}}( \sqrt{\beta}|01\rangle + \sqrt{1 - \alpha - \beta}|10\rangle), \label{eq:gWoutcomes1} \\
&&|\zeta_1\rangle = |00\rangle.
\label{eq:gWoutcomes}
\end{eqnarray}
with $p_0 = 1-\alpha$ and $p_1 = \alpha$. Following the protocol described in Sec. \ref{W}, we can easily find that $f_{max}^{gW}$ can be maximized.
If we now assume that the fidelity of local measurement-based single-path QR is maximized when the fidelities at each segment (i.e., $A\rightarrow C_1$ and $C_1 \rightarrow B$) are maximized, the optimal measurement basis is the computational basis. \hfill $\blacksquare$

After proving the optimality of the measurement basis, let us now  explicitly evaluate the optimal fidelity in this process. 
If \(C_2\) performs measurements  in the computational basis on her parts of the shared states,
 the PMS between Alice (Bob) and $C_1$ are given in Table \ref{T:gW_linear_chain}.
\begin{table}[ht]
\begin{center}
\begin{tabular}{| C{1.7cm} | C{1.9cm} | C{2.7cm} | N}
\hline
\cellcolor{LightRed}{\bf Outcome} & \cellcolor{Yellow}{\bf Probability} & \cellcolor{LightGreen}{\bf Output state}&\\[1ex]
\hline
 $|0\rangle|0\rangle$ & $p_0p_0$ & $|\zeta_0\rangle_{AC_1}|\zeta_0\rangle_{BC_1}$ &\\ [1ex]
\hline
$|0\rangle|1\rangle$ & $p_0p_1$ & $|\zeta_0\rangle_{AC_1}|\zeta_1\rangle_{BC_1}$ &\\ [1ex]
\hline
$|1\rangle|0\rangle$ & $p_1p_0$ & $|\zeta_1\rangle_{AC_1}|\zeta_0\rangle_{BC_1}$ &\\ [1ex]
\hline
$|1\rangle|1\rangle$ & $p_1p_1$ & $|\zeta_1\rangle_{AC_1}|\zeta_1\rangle_{BC_1}$ &\\ [1ex]
\hline
\end{tabular}
\caption{Table of measurement outcomes by $C_2$. The probabilities of getting the outcomes and 
the post measured states, shared between Alice (Bob) and $C_1$, when two copies of three-qubit generalized W
states are shared. The output states, $|\zeta_0\rangle$ and $|\zeta_1\rangle$, are given in Eqs. (\ref{eq:gWoutcomes1})  and (\ref{eq:gWoutcomes}) respectively. }
\label{T:gW_linear_chain}
\end{center}
\end{table} 
  From this table, it is clear that except the first outcome, a pure product
state is shared between $A$ and \(C_1\), or between \(C_1\) and $B$, or between both in all the three cases, and hence for these situations, 
the fidelity reduces to  $\frac 23$.  In the first situation, i.e., when the outcome is $|0\rangle|0\rangle$, the   fidelity  reads as $\frac 23 +  \frac{4}{3 (p_0)^2} \beta (1 - \alpha -\beta)$.
%,  as $|\zeta_0\rangle$ is shared among the 
 For a detailed calculation see \ref{sec:two_linearchain}. Therefore,  the optimal fidelity in the local
measurement-based single-path QR, when the initial shared state is the generalized W state,  can be shown to be
\begin{eqnarray}\label{eq:fidel_line_gW}
{\cal F}^l_2(gW) &=& \left(1 - (p_0)^2\right)  \frac{2}{3} + (p_0)^2 \left( \frac 23 +  \frac{4}{3 (p_0)^2} \beta (1 - \alpha -\beta)\right) \nonumber \\
&=& \frac 23 +  \frac{4}{3} \beta (1 - \alpha -\beta).
\end{eqnarray}
\emph{Remark.} For the shared gW state, instead of  performing any measurement, if \(C_2\) leaves her laboratory, or if  \(C_1\)  tries to communicate the quantum state secretly to Bob ignoring \(C_2\), one can show that  sending  unknown quantum state with fidelity better than the classical is still possible, and the fidelity is given by
\begin{equation}\label{eq:fidel_gW_lin}
{\cal F}^l_{\text{mixed}}(gW) = \frac 23 +  \frac{2}{3} \Big(2\beta (1 - \alpha -\beta) - \alpha(1-\alpha)\Big). 
\end{equation}
%if  the same protocol as discussed above is followed. 
Since,  $\alpha (1-\alpha) > 0 $, we conclude that the local measurement-based protocol is always better than the scheme where \(C_2\)  just leaves the protocol without performing the measurement. 

Let us now state one of the main results of the paper. In particular, we compare the quantum capacities of  multipath QR with that of the local measurement-based single-path ones. 
Before discussing the results for the entire family of the  gW state, let us state the following theorem which is true  for the W state with \(\alpha= \beta = \frac{1}{3}\): \\
{\bf Theorem:} \textit{If two copies of the three-qubit W state are shared between $A,C_1,C_2$ and $C_1,C_2,B$,  sending an unknown qubit in the  teledistribution and teleconcentration protocol is always beneficial than using a single-path quantum repeater, consisting of two bipartite quantum states derived from optimal local 
measurements in the non-functioning port, and the corresponding advantage is $9.1\%$ or better.}

\textit{Proof:} 
For the W state, $\ket{W} = \frac{1}{\sqrt{3}} (\ket{001} + \ket{010} + \ket{100})$, 
%i.e. for $\alpha = \beta = \frac 13$,
${\cal F}^{DC}(W) \geq \frac 89 > \frac{22}{27} 
= {\cal F}^{l}_2(W)$, by using Eqs. (\ref{eq:fidel_gW_TDTC}) and (\ref{eq:fidel_gW_lin}), thereby establishing the advantage of  mutipath protocol for sending quantum information over  the single-path ones. 
% completes the proof. \\ 
\hfill $\blacksquare$

The above Theorem holds even for the shared generalized W states, for some values of $\alpha$ and $\beta$. In particular, we show that the TD-TC scheme is better than the optimal local measurement-based ones 
when $$\alpha \geq \frac{\beta}{2}.$$ In general, for gW states with $\alpha \geq \frac{\beta}{2}$, the TD-TC protocol performs better than the local measurement based scheme by $\frac{(2\alpha-\beta)(1-\alpha-\beta)}{1+2\beta(1-\alpha-\beta)}\times 100\%$ or better. Note, however, that  it is still possible that the region in which the TD-TC protocol can not give better fidelity, may show the benefit over the local measurement-based scheme if one can construct optimal measurements and unitaries involved in the two-path protocol. 

%\begin{center}
% \begin{figure}
% \includegraphics[scale=0.8]{gGHZc4.pdf}
% \caption{\label{noise_m} The noisy case of figure~\ref{noise}, coupled with an optimal measurement performed by Claire 2, 
% resulting in the conversion of the setup into a linear chain.}
% \end{figure}
%\end{center}
% 
 \begin{table*}
\begin{center}
\begin{tabular}{| M{3.5cm}  | M{6.5cm} | M{6cm} | N|}
\hline
\cellcolor{LightGreen}{ } & \multicolumn{2}{|c|}{\cellcolor{LightRed}{\bf Fidelity}}&\\ [1.2ex]
\hline
\cellcolor{LightGreen}{\bf Noise  type} & \cellcolor{LightCyan}{\bf Multipath q. repeater} & \cellcolor{Yellow}{\bf Single path q. repeater}&\\[2.5ex]
\hline
 Bit flip & $\frac{2}{3}+\frac{4}{3}\alpha(1-\alpha)$ & $\frac{2}{3}\{1-p(1-p)\}+\frac{4}{3}\alpha(1-\alpha)\{1-2p(1-p)\}$ &\\ [2ex]
\hline
Phase flip & $\frac{2}{3}+\frac{4}{3}(1-2p)^2\alpha ( 1-\alpha)$ & $\frac{2}{3}+\frac{4}{3}(1-2p)^2\alpha ( 1-\alpha)$ &\\ [2ex]
\hline
Bit phase flip & $\frac{2}{3}+\frac{4}{3}(1-2p)^2\alpha ( 1-\alpha)$ & $\frac{2}{3}+\frac{4}{3}(1-2p)^2\alpha ( 1-\alpha)$ &\\ [2ex]
\hline
Amplitude damping & $\frac{2}{3}+\frac{4}{3}\alpha(1-\alpha) -\frac{2p}{3}(1+\alpha-2\alpha^2-p+2\alpha p-\alpha^2p)$ & $\frac{2}{3}+\frac{4(1-p)}{3}\alpha(1-\alpha)$ &\\ [2ex]
\hline
Phase damping & $\frac{2}{3}+\frac{4}{3}\alpha(1-\alpha) -\frac{4\alpha p}{3}(2 -2\alpha -p +\alpha p)$ & $\frac{2}{3}+\frac{4}{3}\alpha(1-\alpha) -\frac{4\alpha p}{3}(2 -2\alpha -p +\alpha p)$ &\\ [2ex]
\hline
\end{tabular}
\caption{Effects of five kinds of paradigmatic noise model  on fidelities obtained in the TD-TC and local measurement-based schemes when two copies of the three-qubit gGHZ state are shared between input, auxiliary and output nodes. }
\label{T:fidelity_noise}
\end{center}
\end{table*}

\begin{figure}[t]
\begin{center}
 \includegraphics[scale=0.8]{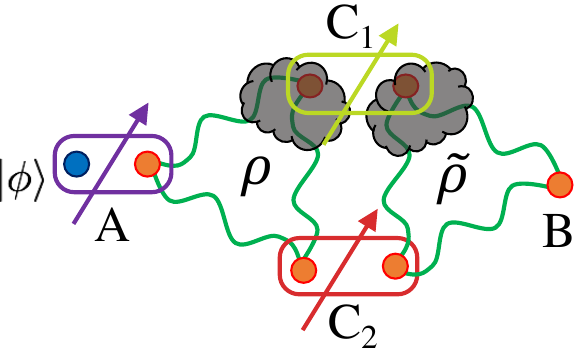}
 \end{center} 
 \caption{\label{noise} Two-path QR where two qubits of \(C_1\) are affected by local noises. In case of local measurement-based single-path scheme, local measurements are performed on both the qubits of \(C_2\) which are not affected by any kind of noise. }
\end{figure}

Note that in the case of local measurement-based single-path quantum repeater, using Bell measurements, we obtain the maximal possible fidelity attainable for  non maximally entangled bipartite states. However, restricting to Bell measurements in the multipath  setting is an assumption in our case. We are prompted to make this assumption due to two main reasons:
\begin{enumerate}
\item Apart from the intrinsic advantage that the multipath protocol offers in terms of security in delaying the teleportation protocol appropriately, we want to compare single-path and multiple path protocols when both of them use the same setup. In this situation, we find wide parameter ranges in both noiseless and noisy scenarios (which we deal in the subsequent section), where the later performed better in terms of fidelity and detection as well as correction of local errors. Since the single- path protocol is the optimal one, after optimization over measurements, the fidelity of the multipath scheme can only improve and hence our conclusion remains same. 
\item If we want to use any other measurement setting, one requires to perform a nonlinear optimization over multiple parameters which in some sense intractable even numerically.
\end{enumerate}

%\textcolor{red}{The above computability argument, coupled with the advantage in terms of fidelity and robustness against certain local noises of such a multipath scheme, even  in its non optimal version prompted us to make the simple assumption of using Bell measurements.}

\section{Noisy channels}\label{sec:noise_tele}
%%%%%%%%%%%%%%%%%%%%%%%%%%%%%%%%%%%%%%%%%%%%%%%%%%%%%%%%%%%%%%%%%%%%%%%%%%%%%%%%%%%%%
Until now, the results obtained  involve quantum channels which are not affected by any kind of noise.
In practice,  quantum states can never be kept completely isolated from the environment, 
and hence it is important to investigate  the effects of environmental interaction on the quantum capacities  of the protocol.
In particular, we assume that local noise acts on both the qubits, possessed by 
$C_1$, as shown in Fig.~\ref{noise} in both multipath and local measurement-based single-path protocols. 
%Claire 2 performs measurements in the optimal basis (see Fig. \ref{noise_m}). 
In this section, we find out the robustness of fidelities, against  noise  in one of the subsystems of the 
shared channels.
% and compare it with the measurement-based single-path quantum repeater
The initial shared state is again  two copies of a three-qubit generalized GHZ or a generalized W state.
We will consider five different kinds of noise \cite{preskil} models  -- 
%\begin{enumerate}
(1) Bit flip noise, (2)  phase flip noise, (3) bit-phase flip noise, (4) amplitude damping, and (5) phase damping.
The detailed actions of these noise models on the quantum state are given in Appendix \ref{sec:noise}.
We  assume that irrespective of the noise acting on the subsystems, Alice, Claires (\(C_1\), and \( C_2\)) and Bob continue with their protocol of the noiseless scenario, described in Secs. \ref{sec:dis_con} and \ref{sec:linearchain}. This is probably a natural and important assumption since we believe that the senders and the receivers may not always be in a position to alter their actions, depending on the noise or they may not always be aware of the types of noise acting on the system.

\begin{table*}[ht]
\begin{center}
\begin{tabular}{| M{3.5cm}  | M{6.5cm} | M{6cm} | N|}
\hline
\cellcolor{LightGreen}{\bf Noise } & \multicolumn{2}{|c|}{\cellcolor{LightRed}{\bf Fidelity}}&\\ [1.2ex]
\hline
\cellcolor{LightGreen}{\bf  type} & \cellcolor{LightCyan}{\bf Multipath QR} & \cellcolor{Yellow}{\bf Single-path QR}&\\[1.2ex]
\hline
 Bit flip & $\frac{2}{3} + \frac{2}{3}\Big((2 \alpha +\beta)(1 - \alpha -\beta) \times  $ $  (1 - 2p +2p^2)+ p(p - 1) \Big)$ & $\frac 23 +\frac 43 \beta(1 - \alpha -\beta) - \frac 23 p(1 - p)\times $ $\left((1 - \alpha)^2-4\beta(1 - \alpha - \beta)\right) $ &\\ [2ex]
\hline
Phase flip & $\frac{2}{3}  + \frac{2}{3}\Big(2\alpha (1 - \alpha) + \beta(1 - \beta)(1-2p)^2$ $  -3\alpha\beta + 4\alpha\beta p(1 - p)\Big)$ & $\frac 23 +\frac 43 \beta(1 - \alpha -\beta)\big(1 - 4p(1-p)\big)$ &\\ [2ex]
\hline
Bit phase flip & $\frac{2}{3} + \frac{2}{3}\Big((2 \alpha +\beta)(1 - \alpha -\beta)  \times  $ $  (1 - 2p +2p^2)+ p(p - 1) \Big)$ & $\frac 23 +\frac 43 \beta(1 - \alpha -\beta) - \frac 23 p(1 - p)\times $ $\left((1 - \alpha)^2-4\beta(1 - \alpha - \beta)\right) $ &\\ [2ex]
\hline
Amplitude damping & $\frac{2}{3} + \frac{2}{3}\Big(  (2 \alpha +\beta)(1 - \alpha -\beta)$ $+ p\big(\alpha\beta -2\beta + \beta^2(1 + p)\big)\Big)$  & $\frac 23 +\frac 43 \beta(1 - \alpha -\beta) - 2\beta p (1 - \alpha) $ $+ \frac 23 p\beta^2(2 + p)$&\\ [2ex]
\hline
Phase damping & $\frac{2}{3}  + \frac{2}{3}\Big(2\alpha (1 - \alpha) + \beta(1 - \beta)(1-p)^2$ $  -3\alpha\beta + \alpha\beta p(2 - p)\Big)$ & $\frac 23 +\frac 43 \beta(1 - \alpha -\beta)\big(1 - p(2-p)\big)$  &\\ [2ex]
\hline
\end{tabular}
\caption{Similar consideration as in  Table \ref{T:fidelity_noise} when the shared state is the generalized W state instead of the gGHZ state. }
\label{T:fidelity_noise_gW}
\end{center}
\end{table*} 
 
\subsection{Generalized GHZ state against noise: Inherent detection and rectification}
%%%%%%%%%%%%%%%%%%%%%%%%%%%%%%%%%%%%%%%%%%%%%%%%%%%%%%%%%%%%%%%%%%%%%%%%%%%%%%%%%%%%%
As before,  $A$ ($B$), $C_1$ and $C_2$ share a $\ket{gGHZ (\alpha)}$ state, and local noise acts on both the subsystems of $C_1$. Before comparing the fidelities obtained from the two-path  and local measurement-based single-path QRs, let us first discuss that the multipath TD-TC protocol enables us to identify and rectify certain kinds of noise in the system when the gGHZ state is shared between them.

%\subsubsection{Detection of Bit flip error}
First note that in the TD-TC scheme, due to the symmetry of the gGHZ states,  
when $\ket{\phi^\pm}(\ket{\psi^\pm})$ clicks in the measurement performed by  $C_1$, the outcome of the measurement  at the node of
$C_2$  can not  be  $\ket{\psi^\pm}(\ket{\phi^\pm})$ in a noiseless scenario or when phase flip error occurs. However, when bit flip or bit-phase flip noise
happens on both the qubits of  $C_1$, such correlations in measurement results are broken and new correlations appear, leading to identification of noise models acting on \(C_1\). 
In particular, in these cases,   when $\ket{\phi^\pm}$ clicks in $C_1$'s port, 
$\ket{\psi^\pm}$ can only be the outcome at $C_2$ and vice-versa. 
 Therefore, this contrasting feature in the measurement outcomes enables us to  conclusively distinguish 
the local bit flip and bit-phase flip  noises acting on one of the Claires' subsystem in the TD-TC 
protocol from the noiseless and phase flip noisy scenarios. Interestingly, by designing suitable local unitary operators at  Bob's port, the effects of noise, either bit or bit-phase flip errors,  on fidelities can be corrected.  %as was done in a noiseless scenario. 
%The only difference is that $I$ is applied 
Let us illustrate the unitary operators of $B$ which are appropriate to correct the bit flip error at the node of \(C_1\). When the measurement outcomes at the end of \(C_1\) and \(C_2\) are
 $\{\ket{\phi^+} \ket{\psi^+}, \ket{\phi^-} \ket{\psi^-}\}$,  $\{\ket{\phi^+} \ket{\psi^-}, \ket{\phi^-} \ket{\psi^+}\}$,  $\{\ket{\psi^+} \ket{\phi^+}, \ket{\psi^-} \ket{\phi^-}\}$ and $\{\ket{\psi^+} \ket{\phi^-}, \ket{\psi^-} \ket{\phi^+}\}$, the corresponding unitary operators at the output port can  be set to  \(\{I, \sigma_z, \sigma_x, \sigma_y\}\) respectively. 
 % $\sigma_z$ has to be applied, when  is the outcome, $\sigma_x$ is for while $\sigma_y$ is for .
If Bob now employs the above set of unitaries, the lower bound on the fidelity for TD-TC  under bit flip channel coincides with the noiseless case. Interestingly, the the detection and subsequent rectification are not possible in the local measurement-based case, where the fidelity depends on the noise parameter, $p$, (see Table \ref{T:fidelity_noise}). This rectification procedure again shows another superior characteristic of multipath QR protocol over the single-path one.

Suppose now that the sender and the receiver a priori know the more probable error at the \(C_1\)'s port to be the bit-phase flip one.  Hence the circuit for implementing the multipath protocol can be designed in such a way that  the fidelity remains unaltered in this scenario. 
%In case of bit phase flip noise, the same kind of complementarity in the  measurement outcome between noiseless and noisy channels in the Claire 1 and 2's ports  can also 
%be observed  and hence enabling detection of an occurance of an error. 
%In other words if Claire 1 and Claire 2 get $\{\ket{\phi^\pm}\}$ and $\{\ket{\psi^\pm}\}$ outcome respectively or vice-versa then they conclude that bit flip or bit-phase flip noise takes place at Claire's port while if outcomes are  $\{\ket{\phi^\pm}\}$ and $\{\ket{\phi^\pm}\}$ or  $\{\ket{\psi^\pm}\}$ and $\{\ket{\psi^\pm}\}$, then either there is no noise or a phase flip noise at Claire 1's part.
%However, unlike the bit flip case,  same unitaries at the Bob's side do not repair the damage occurred in the fidelity as is still depends on the noise parameter (see Table \ref{T:fidelity_noise}).
%However, note that if it is apriori known to Claire 1 that bit-phase flip noisy channel happens, she can tell that to Bob and 
Specifically, the bit-phase flip error can be corrected if Bob  applies $\{\sigma_z\otimes U_i\}$ ($i = 1,2,\ldots, 8$) where $\{U_i\}$ are the set of unitaries used in the bit flip case.
% Similarly, it is not possible to correct the phase flip channel.
 In presence of different kinds of noise models, the  lower bounds on ${\cal F}^{DC}$, and the exact values of ${\cal F}_2^l$ with the measurement being performed in the $\{\ket{+},\ket{-}\}$ basis  are given in Table \ref{T:fidelity_noise} for the shared  gGHZ state. 
As discussed before, for bit flip/bit-phase flip errors, the  multipath setting turns out to be maximally robust compared to any other noise models while  the opposite occurs for the amplitude damping ones, provided the set of measurements and unitaries discussed before in the TD-TC protocol for the shared gGHZ state is optimal. On the other hand, the lower bounds on the TD-TC, match with ${\cal F}_2^l$, when other noises like phase flip,  bit-phase flip,  phase damping act locally on \(C_1\),  if we assume that the rectifying unitary operators are only applied for bit flip errors.

Remark. For an arbitrary noise model, for example for amplitude damping, when Claires perform 
measurements in the Bell basis, all possible (sixteen) outcomes can arise, and hence Bob applies all the sixteen 
unitary rotations mentioned in the noiseless and bit flip cases. 
 In this sense, when the shared state is  the gGHZ state, one can always distinguish amplitude damping or other noise models  from the errors like bit, phase, bit-phase flips and phase damping channels, although rectification might not always be possible.

\subsection*{Effects of local bit-flip noise on the intermediate ports}

In the above considered situation, we have taken the noise to act on only one node $(C_1)$. Since these two nodes ($C_1$ and $C_2$) are at distant (different) locations, we can safely assume that presence or absence of noise in one of the nodes is completely uncorrelated to that of the other. Thus, we argue that it is quite likely that one of the two nodes suffer from local noise, thereby rendering one node to be noisy and the other to be completely error-free. 
%Therefore, we have taken the noise to act on only one node, in our case $(C_1)$, throughout the analysis. Nevertheless, our investigations can be easily extended for tackling noise in both the connecting nodes. 
Let us consider how the situations when both $C_1$ and $C_2$ are affected by local noise. Specifically, we present a detailed analysis of fidelity in the presence of bit-flip noise in both the nodes for the gGHZ state. 
The calculation can be prototypically carried out for other noise models also.

When the shared gGHZ states in both the distribution and concentration sector are affected by noise in both the connecting nodes, the fidelity of the multipath teleportation protocol (designed to correct local bit flip noise in $C_1$) reads as
\begin{eqnarray}
\mathcal{F}^{DC} = \frac{2}{3}\{1- q(1-q)\}+ \frac{4}{3}\alpha(1-\alpha)\{1- 2q(1-q)\}, \nonumber \\
\end{eqnarray}  
Here, $p$ and $q$ are the probabilities with which bit-flip noise acts on $C_1$ and $C_2$ nodes respectively and the noise on $C_!$ can be corrected as discussed before. Note, when $q=1$ or $0$, the expression of fidelity reduces to that of the noiseless scenario. For intermediate values of $q$, the perfect detection and correction of error is not possible. Nevertheless, in the multiparty scenario, we still obtain an error detection probability of $2q(1-q)\{1-2p(1-p)\} + 2p(1-p)\{1-2q(1-q)\}$. 
%$2\big(p^2q(1-q)+pq^2(1-p)+p(1-p)(1-q)^2+(1-p)^2q(1-q)\big)$. 
As expected, for $q=1$, the detection probability is unity for all values of p, while for $q=1/2$, the same falls to $1/2$ for the entire $p$-range. The corresponding linear chain fidelity, where $C_2$ leaves the protocol after performing the measurement, is given by
\begin{eqnarray}
\mathcal{F}^{LC}_{C_2} = \frac{2}{3}\{1- p(1-p)\}+ \frac{4}{3}\alpha(1-\alpha)\{1- 2p(1-p)\}. \nonumber \\ 
\end{eqnarray}
On the other hand, when $C_1$ leaves the protocol, we have 
\begin{eqnarray}
\mathcal{F}^{LC}_{C_1} = \frac{2}{3}\{1- q(1-q)\}+ \frac{4}{3}\alpha(1-\alpha)\{1- 2q(1-q)\}. \nonumber \\
\end{eqnarray}
Note that if the knowledge of which node ($C_1$ or $C_2$) suffers from more noise is available, the DC protocol can be chosen such that the highest fidelity obtained is 
\begin{eqnarray}
\mathcal{F}^{DC} = \max \{\mathcal{F}^{LC}_{C_1}, \mathcal{F}^{LC}_{C_2}\}.
\end{eqnarray} 
Therefore, we can assert that the fidelity obtained in the TD-TC protocol is equal or greater than the linear chain scheme. Here, we have assumed that the party (if any) which leaves the protocol, is typically unknown.
%\underline{In this case, there is no detection of error and the protocol} \underline{can be appropriately chosen such that}
%$\mathcal{F}^{DC}>\mathcal{F}^{LC}$ in the entire range of errors.}

  \subsection{Generalized W state: Quantitative analysis}
  
%Let us now move to the case of when Alice (Bob) and Claire 1 and Claire 2 share the gW state and local noise acts on Claire 1.
We now consider the similar scenario as discussed above, when the initial state is the gW state. 
 Fidelities in two different processes under  different noisy channels are listed in Table \ref{T:fidelity_noise_gW}.
%If the shared channel in our distribution and concentration protocol is a three-qubit generalized W, then the teleportation fidelity for both kind of setup under five different noisy channels is given in Table \ref{T:fidelity_noise_gW}. 
Note that in  case of TD-TC and local measurement-based schemes, both the bit flip and the bit-phase flip errors have similar effects on  fidelities. 
%Similarly, the change in fidelities in presence of local bit flip and bit-phase flip noises are same also in the measurement based scheme.
%However, it is again important to stress that we only have the lower bound on fidelity on TD-TC.
In both the cases, the local measurement-based single-path protocol 
outperforms the two-path  ones for most of the regions in system parameters of gW, i.e., in  $(\alpha, \beta )$-plane.
 On the other hand, there is a contrasting situation in presence of phase damping and phase flip errors  --
TD-TC yields a better fidelity than the single-path case for almost the entire parameter space of gW state. Moreover,   there exists a finite region in the $(\alpha,\beta)$-plane of the shared gW state with local amplitude damping noises at \(C_1\)
where  multipath performs better than the single ones.

To put things in a quantitative perspective, we calculate percentages of gW states for which two-path TD-TC protocol provides better quantum capacities  than that of the local  measurement-based single-path ones under the actions of various local noises. For such comparison, we fix the noise parameter at $p=0.3$, for all the noise models.   We  generate $10^6$ gW states according to the parameterisations, given in \cite{dur2000}.
The observations are the following: 

\begin{enumerate}

\item  For bit flip and bit-phase flip noises, only $9.1\%$ of the states are useful in TD-TC setting. 

\item  In case of amplitude damping channel, the percentage  of the TD-TC with multipath goes up to $41.9\%$. 

\item  For phase damping and phase flip errors, ${\cal F}_2^l (gW) \leq {\cal F}^{DC} (gW)$ holds for $80.3\%$ and $92.6\%$ respectively, thereby establishing the two-path protocol as a better ones than the local measurement-based ones.   

\end{enumerate}

%\subsection{Noise in both the connecting nodes}

\begin{figure}
\begin{center}
\includegraphics[width=0.6\linewidth]{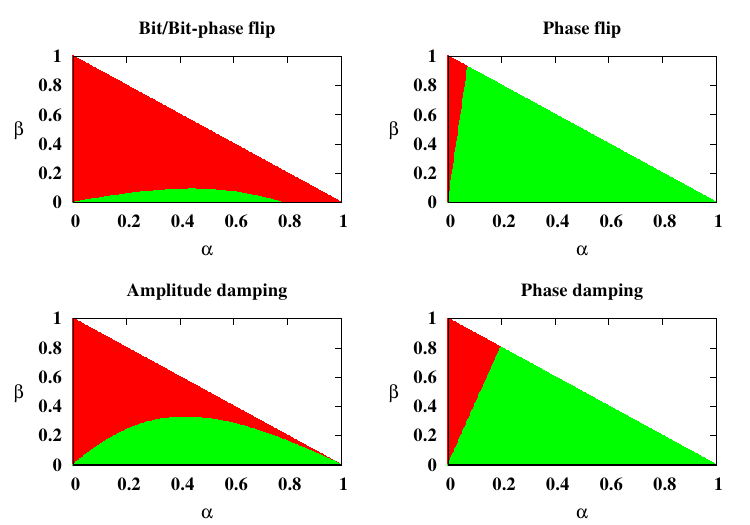}
\end{center}
\caption{The green regions indicate the gW state in the \((\alpha, \beta)\)-plane. In this region,  teleportation via distribution and concentration protocols yields a higher fidelity than the local measurement-based single-path ones which is marked by red. Here $p=0.3$. Both axes are dimensionless.}
\label{fig:error_gw}
\end{figure}

 \begin{figure}[h]
  \begin{center}
 \includegraphics[scale=0.25]{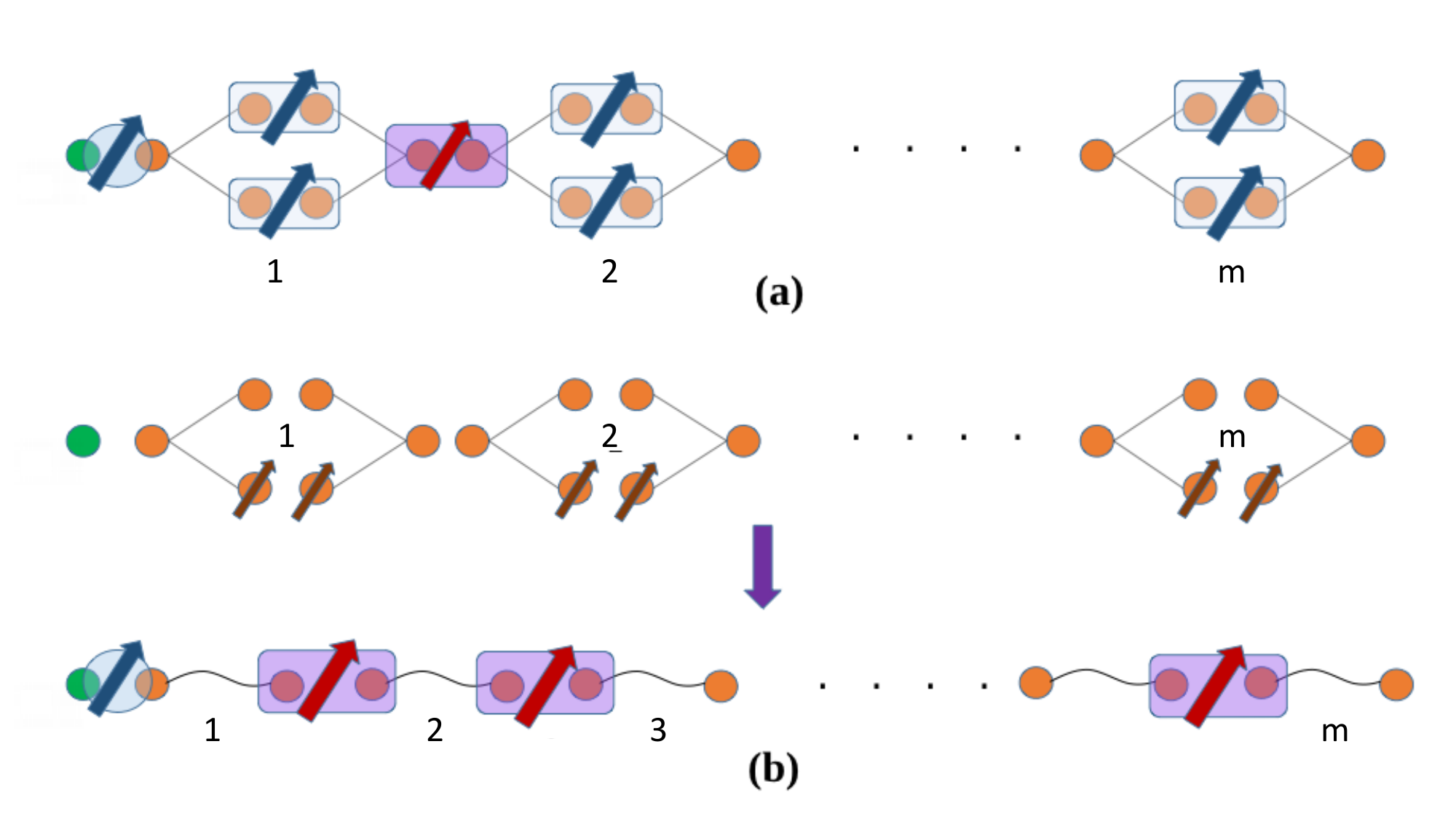}
 \caption{\label{fig:Ncopy_dis_lin} 
(a) Multipath QR: a schematic diagram of QR with \(m\) blocks, each consists  of a multipath TD-TC scenario with shared tripartite entangled quantum states. (b) Local measurement-based QR: in each block,  one of the parties performs an optimal local measurement on her/his qubits, which converts each block to two  single-path quantum channels.}
 \end{center}
\end{figure}

 \section{Multiple blocks of Multipath vs. Local measurement-based Protocols}
 \label{Sec:multi_block}

%Both in noiseless and noisy scenarios, we have already found that multipath QR protocol outperforms the measurement based single path one in many situation when initially a three qubit state is shared. 

%If our aim is to establish  a long distance quantum internet, the protocols discussed above should form a single segment of the entire channel, consisting of several such blocks. 
%The effects of decoherence in a protocol, which can be intentional or unintentional, increase with the distance of the path in which the information should travel.  Hence in quantum state transmission, the division of the entire length into small blocks has high importance, since  input, auxiliary and output nodes then have freedom to abandon the protocol whenever necessary.  
%%As also discussed before, in each step, involvement of all the parties play a crucial role.

We now illustrate the multiple blocks scheme with tripartite states which can be easily generalized to arbitrary number of parties. 
As shown in Fig. \ref{fig:Ncopy_dis_lin}, the total length between the sender and the receiver is divided into 
arbitrary, say $m$, number of  units (blocks) -- each unit shares two copies of the same given state as in Fig. \ref{fig:TDTC}, which execute a multipath TD-TC scheme following the same steps as given in the beginning of Sec. \ref{sec:dis_con}. The output state after implementing TD-TC protocol for $i-1$ blocks, becomes the input state for the $i^{\text{th}}$ block. See Fig. \ref{fig:Ncopy_dis_lin}(a).
%It is of course a first step towards long distance quantum internet, where the sender, Alice and the receiver, Bob is situated in two distant locations.
%Specifically, we assume that  Alice shares two three-party  quantum states with first Bob, $B^1$, then he  is connected with the second Bob,  $B^2$ and so on -- finally $B^{m-1}$ are joined with $B^m$ via two tripartite  quantum channels  where $C^m_1$ and $C^m_2$ are the auxiliary nodes
%%Let us now suppose that there are $m$ such blocks of distribution and 
%%concentration, between Alice and Bob, each distribution and concentration channel, in turn, 
%%composed of N parties,  which is clearly 
%(see Fig. \ref{fig:Ncopy_dis_lin}). At the end, the arbitrary state possessed by the input port, $A$ should reach to the final node, $B^m$ with the help of $2 m$ tripartite quantum channels. 
%The setup behaves like a quantum repeater of $m$ blocks, where each 
%block, once again, consists of $N-1$ distribution and $N-1$ concentration channels. 
%Teleporatation is thus carried out across these $m$ blocks of quantum repeaters with a fidelity
We are interested in evaluating  quantum capacities of an entire quantum channel in terms of the fidelity, ${\cal F}_m^{DC}$, after $m$ blocks.
%of being reproduced at Bob's end.
We will  compare the above scenario with the local measurement-based scheme. In the local measurement-based scheme,  as before, one of the Claire's of each block,  perform optimal local measurements, reducing the entire block structure to a single-path,  consisting  of $2 m$ bipartite states as depicted in Fig. \ref{fig:Ncopy_dis_lin}(b).

When three-qubit gGHZ states are used as quantum channels and Bell measurements as well as the same unitary operations are performed as described before,   the fidelity for the entire QR process reads as
\begin{eqnarray}
F_m(gGHZ)  &=& 2^{2m}   \alpha^{m}(1-\alpha)^{m} \int 2|a|^2|b|^2 d^2 a d^2b\nonumber \\
&& \hspace{-2em} + \sum\limits_{k=0}^{2m} \binom{2m}{k}\alpha^k (1-\alpha)^{2m-k} \int \big( |a|^4 + |b|^4 \big) d^2 a d^2b \nonumber \\
&=& \frac{2}{3}+ \frac{2^{2m}}{3}\big(\alpha(1-\alpha)\big)^{m} \leq {\cal F}^{DC}_m(gGHZ).
\label{eq:Nlinearch}
\end{eqnarray}
%which is a lower bound of ${\cal F}^{DC}_m(gGHZ)$. 
On the other hand,  if one obtains $m$ blocks of single-path  quantum channels (which therefore consists of $2m$ bipartite states) by  optimal local measurements which in this case is \(\{\ket{+}, \ket{-}\}\),  iterative methods leads to the fidelity given by
\begin{eqnarray}
{\cal F}^{l}_m(gGHZ) &=& \frac{2}{3}+ \frac{2^{2m}}{3}\big(\alpha(1-\alpha)\big)^{m}, 
\label{eq:repeatergGHZlinear}
\end{eqnarray}
which happens to coincide with $F_m(gGHZ)$, which in turn is upper bounded by ${\cal F}^{DC}_m(gGHZ).$
\begin{figure}
 \begin{center}
 \includegraphics[scale=0.45]{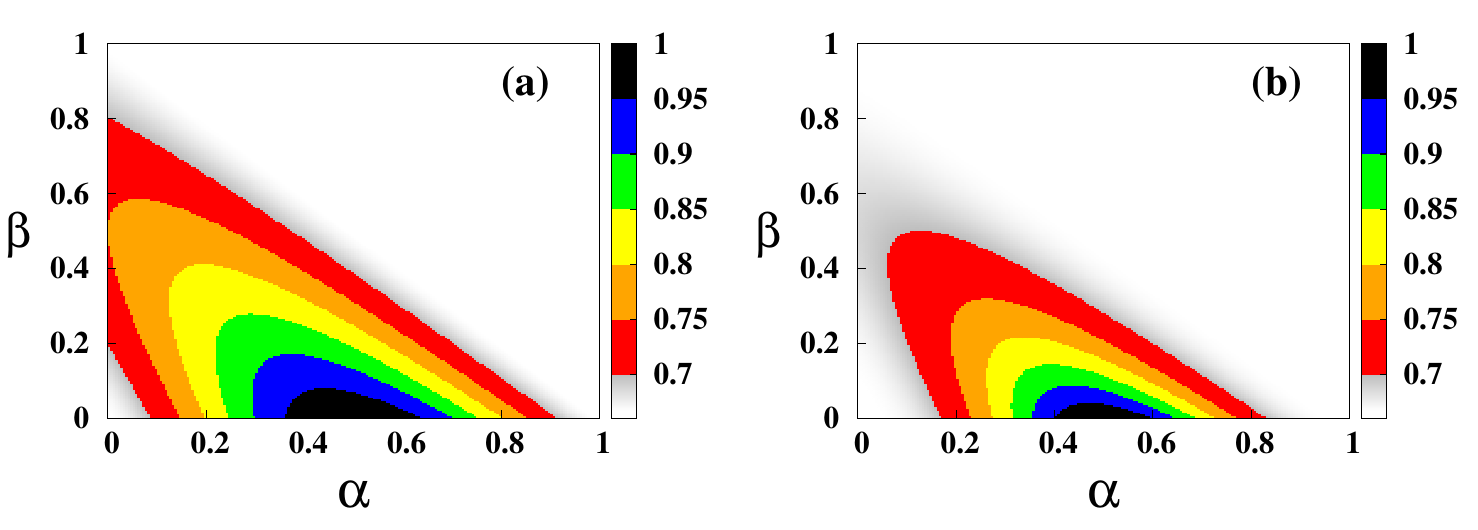}
 \caption{\label{fig:2n4gW} 
Contour plots  of  fidelities for multipath TD-TC scheme with  few blocks in \((\alpha, \beta)\)-plane of the generalized W state.  Number of blocks are chosen as (a)  $m = 2$, (b)  $m = 4$. Both the axes are dimensionless.}
 \end{center}
 \end{figure}
If the similar QR problem is considered for the shared generalized W states of $m$ blocks, the fidelity, after a rather tedious algebraic iterative calculation, is given by
\begin{equation}
F_m(gW) = \frac 23 \Big(1 + 2^{m-1}(2 \alpha +\beta)^m(1 - \alpha -\beta)^m\Big).
\label{eq:gWmblocks}
\end{equation}
%can also be computed with some tedious algebraic iterative calculations while
On the other hand, the local measurement-based protocol leads to the fidelity
\begin{equation}
{\cal F}^{l}_m(gW) = \frac 23 + \frac{2^m}{3}\beta^m(1-\alpha -\beta)^m.
\label{eq:gWmeasurement}
\end{equation}
For comparison, $ F_m(gW)$, for $m = 2$ and $m =4$, is depicted in Fig. \ref{fig:2n4gW}. We observe that the region in the $(\alpha, \beta)$-plane, in which the specific  TD-TC protocol shows quantum advantage, shrinks with the increase of the number of blocks. Moreover, comparing  Eqs. (\ref{eq:gWmblocks}) and (\ref{eq:gWmeasurement}), we find that the benefit of the two-path scheme over the single-path ones by using gW states reduces with $m$. 

\section{Multiple Claires in a single block of the TD-TC scheme} 
Instead of considering multiple distribution-concentration blocks which leads to a repeater like situation, in this section, we consider the effect of multiple Claires (say, $n$ of them) in a single distribution-concentration block. Specifically, if we construct the distribution-concentration block using a $(n+1)$-qubit gGHZ state, which contains $n$ Claires, we find that  the obtained teleportation fidelity is independent of $n$. This is because, due to the symmetry of the gGHZ states, the final states with Bob and their corresponding probabilities remain unaltered with changing $n$. 
In particular, for Bell measurements performed at $n$ Claires, in principle, one expects, $4^n$ clicking combinations. However, as  also noticed in the two-Claire situation, for gGHZ states, the clicks occur with non-zero probability iff all the outcomes are either from the $\phi$-group or from the $\psi$-group. Therefore, for $n$ Claires, one only needs to consider $2^n ~(\phi \text{ group}) + 2^n ~(\psi \text{ group}) = 2^{n+1}$ outcome combinations. Moreover, the post-measurement states for any $\phi (\psi)$ group clickings can be respectively mapped to states that were obtained in the two-Claire scenario (see Table. \ref{T:outcome_ghz}) via local unitary operations, thereby making the fidelity independent of the number of Claires.
This further suggests, that the our protocol for the gGHZ states are optimal resulting the best teledistribution and teleconcentration possible for a gGHZ state with a given value of $\alpha$, which inturn measures its genuine multiparty entanglement content \cite{GGM, GGM2}.
We now want to contrast the two incremental settings considered in the manuscript:
\begin{enumerate}
\item Increasing the number of distribution-concentration blocks $(m)$, leading to a repeater like scenario, and 
\item Increasing the number of Claires $(n)$ in a distribution-concetration block.
\end{enumerate}
Our investigations show that gGHZ states respond in qualitatively  different manner to these increments. In particular, one gets reduction of fidelity on increasing $m$ as shown in Eq. \eqref{eq:Nlinearch}, while it is robust to increasing $n$, as argued above. The robustness of fidelity on increasing $n$ implies that in principle, there is no upper bound to the number of intermediate parties (Claires) that one may employ in a given distribution-concentration block.
Recall, one of our motivations to introduce multiple Claires was to guarantee that the teleportation protocol gets completed at  the stipulated time. 
If the intermediate party is not trustworthy, she/he might teleport the state before the predecided time. To reduce this possibility, Alice decides to resort to multiple Claires, so that the process is completed if and only if all the mediators complete their actions. Therefore, even if some Claires break the trust, Bob does not receive the state before the predecided time.  
Let us elaborate this with an example. Suppose the probability that each of the Claires break trust, i.e, completes the protocol before the stipulated time is $p$. Then for $n$ Claires, the probability that the protocol gets completed before the required time is $p^n$, which approaches zero for large $n$. Therefore, the correct timing of the protocol can be ensured by involving more Claires as per our wish without hampering the fidelity at all.

In addition the above result right away implies that when one considers a repeater-like scenario, it yields the same fidelity as in Eq. \eqref{eq:Nlinearch} since the output states and their probabilities after a distribution-concentration block remain unchanged for any $n$.   Our analysis further reveals that the linear chain obtained by considering $(n-1)$ local measurements, each performed in the $\lbrace |+\rangle, |-\rangle  \rbrace$ bases as seen in Eq. \eqref{eq:fmax}, results in the same two-party state as obtained in the two-Claire scenario. Therefore, the fidelity for the single-path reduced linear chain setting also  remain unaltered, and is given in Eq. \eqref{eq:repeatergGHZlinear}.

In the noisy case, when only one of the $n$ Claires suffer from bit flip, phase flip, or bit phase flip noise, the fidelities again remain insensitive to $n$. Therefore, features of detection, rectification or both of these noises remain true even when the number of intermediate Claires increases. These computations can be exactly carried out like in the noiseless case, noting that the clicking sectors of bit flip and bit phase flip noises are orthogonal to the noiseless scenario, while phase flip noise share the same clicking sector as the noiseless gGHZ state. Therefore, bit flip and bit phase flip errors are auto-detected. Now  one can design the protocol to correct either bit flip or bit phase flip errors. Lastly, we mention that the $n$-Claire-scenario can only be tackled for the gGHZ state and its noisy variants, for which symmetries simplify the calculations.

\section{Conclusion}
\label{sec:conclu}

Quantum teleportation is a pioneering discovery which forms one of the pillars in the success story of quantum information science. Its exclusivity to the quantum domain puts it 
in sharp contrast with classical ideas. In this work, 
%we address the question of sending a quantum state over a long distance in a  multiparty framework.
 we have designed a multiparty teleportation protocol which offers better fidelities compared to single path schemes for some ranges of state parameters both in the noiseless and noisy scenarios. In addition to this, we have also argued that such multiparty schemes (involving distribution and concentration) can be useful in delayed teleportation which  necessitates inclusion of multiple paths and intermediate parties. Specifically, we have considered a teleportation protocol in which the teleporter wants the teleportation process
to be completed at some later, but fixed time, when she/he would not be available in her/his port. So, the teleportation
must be completed by 
intermediate parties who would complete the process at the predecided time. However, to assure the perfect timing of the protocol, one opts for multiple intermediaries, so that even if some of them attempt a premature completion, the process gets accomplished only when all the intermediate parties complete their actions. We intended to achieve this using two entangled multiparty states, where one is used to ``teledistribute" the information, while the other ``concentrates" it back at the predecided time.
In particular, we have
investigated the performance of this multiparty protocol (in terms of its fidelity) for two important families of three-qubit
states, the generalized Greenberger-Horne-Zeilinger (gGHZ) and the generalized W states. We have then compared them with a
protocol consisting of 
linear chains obtained from these multiparty states by performing optimal local measurements in one of the nodes. 
In both noiseless and noisy scenarios, we have shown that for  certain families of multiparty shared states, one shot capacities,
in terms of average output to input fidelity, of multipath protocols are strictly higher than that of the corresponding 
single-path cases. We have also observed that the protocol proposed in this paper inherently possesses a noise correcting
mechanism, when local noise is either bit flip or bit-phase flip, acting on one of the parties of the shared gGHZ state.
%After investigating quantum capacities of two different transmission channels, both bipartite as well as multipartite paths,
Moreover,  we have found the capacities of long distance quantum channels, consisting of arbitrary number of multipartite
as well as bipartite units by using iterative methods. 
Advantages in quantum state transfer  by using multiple path protocol show the importance of creating multipartite
entangled states in quantum communication protocols over the bipartite ones. 

\vspace{0.5cm}

% We finally consider a repeater setup consisting of $m$ distribution concentration blocks and compute the fidelities for both the gGHZ and gW states. To summarise, this work highlights the advantages of teleporting  a qubit using a multiparty teleportation scheme in both noiseless and noisy scenarios. 

\textbf{Acknowledgement:}
This research was supported in part by the `INFOSYS scholarship for senior students'. 

\vspace{0.5cm}
%\end{acknowledgements}

%\vfill
%\break

\appendix

\section{Fidelity of a configuration of two different linear chains}\label{sec:two_linearchain}

Consider now a teleportation protocol with  
 two different linear chains, $|\psi_1\rangle_{AC_1} =\sqrt{\alpha_1} |00\rangle +\sqrt{1-\alpha_1}|11\rangle$,
 connecting Alice and $C_1$ and 
 $|\psi_2\rangle_{C_1B} =\sqrt{\alpha_2} |00\rangle +\sqrt{1-\alpha_2}|11\rangle$,
 connecting $C_1$ and Bob. For such a situation, the optimal teleporation fidelity from Alice to
 Claire (Claire to Bob) can be obtained by performing Bell  measurements and unitary rotations as 
 given in Ref. \cite{bennett1993}. To compute the optimal fidelity for two linear chains, we again use 
 Bell measurements and the respective unitaries.

Suppose, $|\psi\rangle$, is  the arbitrary state that Alice wants to teleport to Bob. In the first step, 
Alice performs a joint Bell measurement on the unknown state and on her part of the shared state, followed 
by a  classical communication of measurement results to Claire. Claire then applies local unitary operator on her part, depending
on the measurement outcomes that Alice obtained. The teleported state (un-normalized) on Claire's part in the first chain, is given in Table \ref{T:protocol1}.
% are  the usual Bloch sphere parameterisations.
\begin{table}[ph]
\begin{center}
\begin{tabular}{|c|c|N|}
\hline
\cellcolor{LightGreen}{Outcome (Alice's)} & \cellcolor{LightRed}{Teleported state} &\\ [1ex]
\hline
$|\phi^\pm\rangle$ & $\frac{1}{\sqrt{2}}\big(a\sqrt{\alpha_1}|0\rangle + b\sqrt{1-\alpha_1}|1\rangle\big)$ &\\ [1ex]
\hline
$|\psi^\pm\rangle$ & $\frac{1}{\sqrt{2}}\big(a\sqrt{1-\alpha_1}|0\rangle + b\sqrt{\alpha_1}|1\rangle\big)$ &\\ [1ex]
\hline
\end{tabular}
\end{center}
\caption{Un-normalized teleported state in the Claire's port, after Alice performs Bell measurements and Claire rotates her part with proper unitary operators.}
\label{T:protocol1}
\end{table}

In the second step, Claire teleports each of the quantum states that she obtains
(see table. \ref{T:protocol}) to Bob by applying the similar protocol as above.

\begin{table}[ht]
\begin{center}
\begin{tabular}{|c|c|N|}
\hline
\cellcolor{LightGreen}{ Outcome (Claire's)} & \cellcolor{LightRed}{Teleported state }&\\ [1ex]
\hline
\multicolumn{2}{|c|}{ \cellcolor{LightCyan}{For $\frac{1}{\sqrt{2}}\big(a\sqrt{\alpha_1}|0\rangle + b\sqrt{1-\alpha_1}|1\rangle\big)$}} &\\ [1.5ex]
\hline
$|\phi^\pm\rangle$ & $\frac{1}{2}\big(a\sqrt{\alpha_1\alpha_2}|0\rangle + b\sqrt{(1-\alpha_1)(1-\alpha_2)}|1\rangle\big)$ &\\ [1ex]
\hline
$|\psi^\pm\rangle$ & $\frac{1}{2}\big(a\sqrt{\alpha_1(1-\alpha_2)}|0\rangle + b\sqrt{(1-\alpha_1)\alpha_2}|1\rangle\big)$ &\\ [1ex]
\hline
\multicolumn{2}{|c|}{ \cellcolor{LightCyan}{For $\frac{1}{\sqrt{2}}\big(a\sqrt{1-\alpha_1}|0\rangle + b\sqrt{\alpha_1}|1\rangle\big)$}} &\\ [1.5ex]
\hline
$|\phi^\pm\rangle$ & $\frac{1}{2}\big(a\sqrt{(1-\alpha_1)\alpha_2}|0\rangle + b\sqrt{\alpha_1(1-\alpha_2)}|1\rangle\big)$ &\\ [1ex]
\hline
$|\psi^\pm\rangle$ & $\frac{1}{2}\big(a\sqrt{(1-\alpha_1)(1-\alpha_2)}|0\rangle + b\sqrt{\alpha_1\alpha_2}|1\rangle\big)$ &\\ [1ex]
\hline
\end{tabular}
\end{center}
\caption{Un-normalized teleported state in Bob's port, depending on Claire's measurement outcomes.}
\label{T:protocol}
\end{table}

The repetitive iteration of the above protocol for the two linear chains, leads to the total fidelity, given by

\begin{eqnarray}
F &=& 4  \int d^2 a d^2b  \Bigg[ \frac{1}{4} \Big[\Big(|a |^2 \sqrt{\alpha_1\alpha_2}+
|b |^2 \sqrt{(1-  \alpha_1)(1 - \alpha_2)} \Big)^2 + \Big(|a |^2 \sqrt{\alpha_1(1 -\alpha_2)}+
|b |^2 \sqrt{(1-  \alpha_1) \alpha_2} \Big)^2  \nonumber \\ &&
\hspace{3cm}+ \Big(|a |^2 \sqrt{( 1 -\alpha_1)\alpha_2}+|b |^2 \sqrt{\alpha_1 ( 1 -\alpha_2)} \Big)^2 + \Big(|a |^2 \sqrt{(1 - \alpha_1)\alpha_2}+ |b |^2 \sqrt{\alpha_1(1- \alpha_2)} \Big)^2\Bigg] \nonumber \\
&=& \frac{2}{3}+\frac{4}{3}\sqrt{\alpha_1\alpha_2(1-\alpha_1)(1-\alpha_2)}.
\label{eq:2linearch}
\end{eqnarray}
Note, when $\alpha_1=\alpha_2=\alpha$, Eq. \eqref{eq:2linearch} reduces to
\begin{eqnarray}\label{eq:linear_fidel}
F = \frac{2}{3}+\frac{4}{3}\alpha(1-\alpha).
\end{eqnarray}

\section{Various noisy channels}\label{sec:noise}
Let us briefly briefly discuss about various kinds of noisy channels \cite{preskil, Nielsen_chuang}
required in the main text. \\
{\bf Bit flip channel:} The bit-flip operation is achieved by applying the Pauli operator $\sigma_x$. 
As the name  suggests, it flips the quantum state $\ket{0}$ to $\ket{1}$ and vice-versa
%The form of bit flip noise we have considered, effects the aforesaid transformation 
with a probability $1 - p$ while it keeps the state unchanged with a probability $p$ , ($0 < p < 1$). 
Hence in the presence of bit flip channel, a quantum state $\rho$ is transformed as 
\begin{equation} 
 \rho \xrightarrow{\text{bit flip}} p \rho+(1-p)\sigma_x\rho \sigma_x.
\end{equation}

{\bf Phase flip channel:} The phase flip operation transforms $\ket{1} \rightarrow -\ket{1}$
%and keeps $\ket{0}$ unchanged. In the phase flip channel used in the paper, nothing happens
%to a quantum state with probability $p$, and the operation occurs with a probability $1-p$.
%In short $\rho$, changes under the presence of phase flip channel as
The transformation in this case reads as
\begin{equation} 
 \rho \xrightarrow{\text{phase flip}} p \rho+(1-p)\sigma_z\rho \sigma_z.
\end{equation}

{\bf Bit-phase flip channel:} Bit-phase flip channel changes a quantum state as follows:
\begin{equation} 
 \rho \xrightarrow{\text{bit-phase flip}} p \rho+(1-p)\sigma_y\rho \sigma_y.
\end{equation}

{\bf Amplitude damping channel:} A quantum state, $\rho$, under the action of amplitude damping 
channel, transforms as
\begin{equation} 
 \rho \xrightarrow[\text{damping}]{\text{amplitude}} M_0 \rho M_0^\dagger + M_1\rho M_1^\dagger,
\end{equation}
where $M_i, ~i = 0,1$, are the Krauss operators, given by
$$
M_0=\left(
\begin{array}{cc}
1 & 0 \\
0 & \sqrt{1 - p}
\end{array}\right)
,~~~~~~~~~~
M_1=\left(
\begin{array}{cc}
0 & \sqrt{p} \\
0 & 0
\end{array}\right),
$$
Satisfying the condition $M_0^\dagger M_0 + M_1^\dagger M_1 = I$.

{\bf Phase damping channel:} The Kraus operator representation of the phase damping channel, when a quantum state $\rho$ is passing through it is given by
 \begin{equation} 
 \rho \xrightarrow[\text{damping}]{\text{phase}} M_0 \rho M_0^\dagger + M_1\rho M_1^\dagger  + M_2\rho M_2^\dagger,
\end{equation}
where 
$$
M_0=\sqrt{1 - p} \left(
\begin{array}{cc}
1 & 0 \\
0 & 1
\end{array}\right)
,~~
M_1= \sqrt{p}\left(
\begin{array}{cc}
1 & 0\\
0 & 0
\end{array}\right)
,~~
M_2= \sqrt{p}\left(
\begin{array}{cc}
0 & 0\\
0 & 1
\end{array}\right).
$$
\pagebreak

\section{Tables}\label{sec:outcome_gW_dis_conc}

\begin{table}[ph]
\begin{center}
\resizebox{\textwidth}{!}{
\begin{tabular}%{|c| M{1.2cm}|c|c|c|c|c|c|}
{ | M{1.2cm} | M{1.2cm}| M{1.3cm} |  M{4.6cm} | M{1.2cm} | M{1.2cm}| M{1.3cm} |  M{4.6cm} | N} 
\hline
  \multicolumn{4}{|c|}{ \cellcolor{LightCyan}{Alice's measurement outcome $|\phi^\pm\rangle$}} & \multicolumn{4}{|c|}{ \cellcolor{LightCyan}{Alice's measurement outcome $|\psi^\pm\rangle$}} &\\ [2.5ex] 
\hline 
  \multicolumn{2}{|c|}{  Outcomes} & Unitary &Fidelity & \multicolumn{2}{|c|}{  Outcomes} & Unitary &Fidelity &\\ [2.5ex] 
\hline 
 $C_1$ & $C_2$ & B & $|\langle \text{out}|\phi\rangle|^2 \times 8$ &$C_1$ & $C_2$ & B & $|\langle \text{out}|\phi\rangle|^2 \times 8$ &\\ [2.5ex] 
\hline

\hline
 $ |\phi^+\rangle$ & $ |\phi^+\rangle$ &   $ I$ &  $ \Big(|a |^2 (\alpha + \beta) +
|b |^2 (1-  \alpha - \beta) \Big)^2$ & $ |\phi^+\rangle$ & $ |\phi^+\rangle$ & $ \sigma_x$ &  $ \Big(|a |^2 (1-  \alpha - \beta) + |b |^2 (\alpha + \beta) \Big)^2$ &\\ [2.5ex] 
\hline
 $ |\phi^+\rangle$ & $ |\phi^-\rangle$ &   $ \sigma_z$ &  $ \big(|a|^2 (\alpha - \beta) +
|b |^2 (1-  \alpha - \beta) \Big)^2$ &$ |\phi^+\rangle$ & $ |\phi^-\rangle$ & $ \sigma_y$ &  $ \Big(|a |^2 (1-  \alpha - \beta) + |b |^2 (\alpha - \beta)\Big)^2 $ &\\ [2.5ex] 
\hline
 $ |\phi^+\rangle$ & $ |\psi^+\rangle$ &   $ \sigma_x$& $\alpha(1 - \alpha - \beta)$  & $ |\phi^+\rangle$ & $ |\psi^+\rangle$ & $ I$& $\alpha(1 - \alpha - \beta)$ &\\ [2.5ex] 
\hline
 $ |\phi^+\rangle$ & $ |\psi^-\rangle$ &   $\sigma_y$ & $\alpha(1 - \alpha - \beta)$ & $ |\phi^+\rangle$ & $ |\psi^-\rangle$ & $ \sigma_z$& $\alpha(1 - \alpha - \beta)$&\\ [2.5ex] 
 \hline
 
 \hline
 $ |\phi^-\rangle$ & $ |\phi^+\rangle$ &   $I$ & $ \Big(|a|^2 (\alpha - \beta) +
|b |^2 (1-  \alpha - \beta) \Big)^2$ & $ |\phi^-\rangle$ & $ |\phi^+\rangle$ & $ \sigma_x$& $\Big(|a|^2 (1-  \alpha - \beta) + |b |^2 (\alpha - \beta) \Big)^2 $ &\\ [2.5ex] 
\hline
 $ |\phi^-\rangle$ & $ |\phi^-\rangle$ &   $ \sigma_z$& $\Big(|a|^2 (\alpha + \beta) +
|b|^2 (1-  \alpha - \beta) \Big)^2$ & $ |\phi^-\rangle$ & $ |\phi^-\rangle$ & $I$ & $\Big(|a |^2 (1-  \alpha - \beta) + |b |^2 (\alpha + \beta) \Big)^2$ & \\ [2.5ex] 
\hline
 $ |\phi^-\rangle$ & $ |\psi^+\rangle$ &   $ \sigma_x $& $\alpha(1 - \alpha - \beta)$ & $ |\phi^-\rangle$ & $ |\psi^+\rangle$ & $ I $& $\alpha(1 - \alpha - \beta)$ & \\ [2.5ex] 
\hline
 $ |\phi^-\rangle$ & $ |\psi^-\rangle$ &  $ \sigma_y$& $\alpha(1 - \alpha - \beta)$ & $ |\phi^-\rangle$ & $ |\psi^-\rangle$ & $ \sigma_z$& $\alpha(1 - \alpha - \beta)$ &\\ [2.5ex] 
 \hline
 
 \hline
  $ |\psi^+\rangle$ & $ |\phi^+\rangle$ &   $\sigma_x$ & $\beta(1 - \alpha - \beta)$ & $ |\psi^+\rangle$ & $ |\phi^+\rangle$ & $ I$& $\beta(1 - \alpha - \beta)$ &\\ [2.5ex] 
 \hline
  $ |\psi^+\rangle$ & $ |\phi^-\rangle$ &   $ \sigma_x$& $\beta(1 - \alpha - \beta)$ & $ |\psi^+\rangle$ & $ |\phi^-\rangle$ & $ I$& $\beta(1 - \alpha - \beta)$ &\\ [2.5ex] 
 \hline
  $ |\psi^+\rangle$ & $ |\psi^+\rangle$ &   $ I$& $ 4 \alpha\beta |a|^4$ & $ |\psi^+\rangle$ & $ |\psi^+\rangle$ & $ \sigma_x$& $4 \alpha\beta |b|^4$  & \\ [2.5ex] 
 \hline
  $ |\psi^+\rangle$ & $ |\psi^-\rangle$ &   $ I $& $0$ & $ |\psi^+\rangle$ & $ |\psi^-\rangle$ & $ \sigma_x $& $0$ &\\ [2.5ex] 
 \hline

 \hline
  $ |\psi^-\rangle$ & $ |\phi^+\rangle$ &   $ \sigma_y$& $\beta(1 - \alpha - \beta)$ & $ |\psi^-\rangle$ & $ |\phi^+\rangle$ & $ \sigma_z$& $\beta(1 - \alpha - \beta)$ &\\ [2.5ex] 
 \hline
  $ |\psi^-\rangle$ & $ |\phi^-\rangle$ &   $ \sigma_y$& $\beta(1 - \alpha - \beta)$ & $ |\psi^-\rangle$ & $ |\phi^+\rangle$ & $ \sigma_z$& $\beta(1 - \alpha - \beta)$ &\\ [2.5ex] 
 \hline
  $ |\psi^-\rangle$ & $ |\psi^+\rangle$ &   $ I $& $0$ & $ |\psi^-\rangle$ & $ |\psi^+\rangle$ & $ \sigma_x$& $0$ & \\ [2.5ex] 
 \hline
  $ |\psi^-\rangle$ & $ |\psi^-\rangle$ &   $ I $& $4 \alpha\beta |a|^4$ & $ |\psi^-\rangle$ & $ |\psi^-\rangle$ & $ \sigma_x$& $4 \alpha\beta |b|^4$ & \\ [2.5ex] 
 \hline
\end{tabular}}
\end{center}
\caption {When the initial state is the generalized W states, the computation of fidelity for each measurement outcomes obtained in the ports of Alice, $C_1$ and $C_2$ are listed. The corresponding unitary operators at Bob's node are also mentioned.}
% Table of Bell measurement outcomes of Claire 1 and Claire 2 
%with the corresponding unitary operations performed by Bob and performance quantified in terms
%of fidelity. The distribution and concentration channels used are noiseless generalized W states.}
\label{Table:secondmeasure}
\end{table}

%It is worth noticing from Table ~\ref{Table:secondmeasure} that for both the PMS states 
%$|\zeta_{\phi}^+\rangle_{C_1C_2}$ and $|\zeta_{\psi}^+\rangle_{C_1C_2}$, when Claires perform the measurements,
%a complementarity of measurement outcomes occur between $\ket{\psi^+}$ and $\ket{\psi^-}$.
%When $\ket{\psi^+}$ clicks for one of the Claires, $\ket{\psi^-}$ never clicks for the other
%and vice-versa. Which is almost similar to the shared generalized GHZ state,
%where the complementarity was much more involved and between two different groups, 
%the $|\phi^\pm\rangle$ and $|\psi^\pm\rangle$ groups of measurement outcomes. 

We notice from the table that like the gGHZ states, when $\ket{\psi^+}$ clicks in one of the Claire's port, say, $C_1$, $C_2$ can never obtain $\ket{\psi^-}$ as her measurement outcome. Note, however that the correlations in the measurement outcomes for gGHZ is more prominent than that of the gW states.


\begin{thebibliography}{99}
\bibitem{Bell-basis} The Bell basis consists of four maximally entangled states, given by
% \begin{eqnarray}
 \(|\phi^{\pm}\rangle = \frac{1}{\sqrt{2}}(|00\rangle \pm |11\rangle)\),
 %, \nonumber \\
 %\text{and}~
 and
 \(|\psi^{\pm}\rangle = \frac{1}{\sqrt{2}}(|01\rangle \pm |10\rangle)\),
 %\nonumber
% \end{eqnarray} 
 where $\{\ket{0},\ket{1}\}$ is the computational basis.
\bibitem{GGM2} The genuine multiparty entanglement $(\mathcal{G})$ content of a $n$-qubit gGHZ $(\sqrt{\alpha}|00...0\rangle + \sqrt{1-\alpha}|11...1\rangle)$ state as measured by GGM \cite{GGM} is computed to be
\begin{eqnarray}
\mathcal{G}(gGHZ) = \left\{
 \begin{array}{cc}
 \alpha & \alpha \leq \frac{1}{2}  \\
 1-\alpha & \alpha > \frac{1}{2}
\end{array}\right..
\end{eqnarray}
\end{thebibliography}
\end{document}